\documentclass[a4,fleqn,usenatbib]{mnras}
\usepackage{etoolbox}
\makeatletter
\patchcmd\@combinedblfloats{\box\@outputbox}{\unvbox\@outputbox}{}{%
   \errmessage{\noexpand\@combinedblfloats could not be patched}%
}
 \makeatother

\usepackage[T1]{fontenc} 
\usepackage{ae,aecompl}

\usepackage{graphicx}	
\usepackage{amsmath}	
\usepackage{amssymb}	
\usepackage{stmaryrd}
\usepackage{bm}
\usepackage{times}
\usepackage{amsmath}
\usepackage{amssymb}
\usepackage{textcomp}
\usepackage{verbatim}

\DeclareMathOperator{\sech}{sech}

\newcommand {\hi} {\ifmmode \ion{H}{I} \else $\ion{H}{I}$ \fi}
\newcommand {\hii} {\ifmmode \ion{H}{II} \else $\ion{H}{II}$ \fi}
\newcommand {\hei} {\ifmmode \ion{He}{I} \else $\ion{He}{I}$ \fi}
\newcommand {\heii} {\ifmmode \ion{He}{II} \else $\ion{He}{II}$ \fi}
\newcommand {\heiii} {\ifmmode \ion{He}{III} \else $\ion{He}{III}$ \fi}
\newcommand {\ts}{\textsubscript}

\title[Efficacy of early stellar feedback]{Efficacy of early stellar feedback in low gas surface density environments} 

\author[R. Kannan et al.]{Rahul Kannan$^{1}$\thanks{E-mail: rahul.kannan@cfa.harvard.edu}\thanks{Einstein Fellow},
Federico Marinacci$^1$,
Christine M. Simpson$^{2,3,4}$, 
Simon C. O. Glover$^5$ 
\newauthor{and Lars Hernquist$^1$}\\
\\
$^{1}$Harvard-Smithsonian Center for Astrophysics, 60 Garden Street, Cambridge 02138, MA, USA\\
$^{2}$Heidelberg Institute for Theoretical Studies, Schloss- Wolfsbrunnenweg 35, D-69118 Heidelberg, Germany\\
$^{3}$Enrico Fermi Institute, University of Chicago, Chicago, IL 60637, USA\\
$^{4}$Department of Astronomy $\&$ Astrophysics, University of Chicago, Chicago, IL 60637, USA\\
$^{5}$Universi\"at Heidelberg, Zentrum f\"ur Astronomie, Institut f\"ur theoretische Astrophysik, Albert-Ueberle-Str. 2, 69120 Heidelberg, Germany
}

\date{Accepted XXX. Received YYY; in original form ZZZ}

\pubyear{2018}

\begin{document}
\label{firstpage}
\pagerange{\pageref{firstpage}--\pageref{lastpage}}
\maketitle

\begin{abstract}
 We present a suite of high resolution radiation hydrodynamic simulations of a small patch ($1 \ {\rm kpc}^2$) of the inter-stellar medium (ISM) performed with {\sc Arepo-RT}, with the aim to quantify the efficacy of various feedback processes like supernovae explosions (SNe), photoheating and radiation pressure in low gas surface density galaxies ($\Sigma_{\rm gas} \simeq 10 \ {\rm M}_\odot \ {\rm pc}^{-2}$). We show that radiation fields decrease the star formation rate and therefore the total stellar mass formed by a factor of $\sim 2$. This increases the gas depletion timescale and brings the simulated Kennicutt-Schmidt relation closer to the observational estimates. Radiation feedback coupled with SNe is more efficient at driving outflows with the mass and energy loading increasing by a factor of $\sim 10$. This increase is mainly driven by the additional entrainment of medium density ({$10^{-2} \leq n< 1  \ {\rm cm}^{-3}$}), warm ({$300 \ {\rm K}\leq T<8000 \ {\rm K}$}) material.  Therefore including radiation fields tends to launch colder, denser and higher mass and energy loaded outflows. This is because photoheating of the high density gas around a newly formed star over-pressurises the region,  causing it to expand. This reduces the ambient density in which the SNe explode by a factor of $10-100$ which in turn increases their momentum output by a factor of $\sim 1.5-2.5$. Finally, we note that in these low gas surface density environments,  radiation fields primarily impact the ISM via photoheating and radiation pressure has only a minimal role in regulating star formation.
\end{abstract}


\section{Introduction}
Stellar feedback has been invoked by many models to explain the low efficiency of star formation in low mass (${\rm M}_{\rm halo} \lesssim 10^{12} {\rm M}_\odot$) haloes \citep{Navarro1996, Springel2003, Stinson2006, DV2008, Agertz2013, Vogelsberger2013, Hopkins2014, Hopkins2017}.   Early cosmological simulations invoked the energy injection from SNe events as the main feedback mechanism to regulate star formation. However, the lack of resolution meant that the coupling between supernovae
(SNe) feedback energy and the ISM was very inefficient \citep{Katz1996,
Navarro1996}, as most of the injected energy is radiated away quite quickly.  This precipitated the need to treat SNe feedback on sub-grid scales, which involved models such as delayed cooling of gas heated by a SNe event
\citep{Thacker2001, Stinson2006, Agertz2013}, stochastic heating of the gas to temperatures where cooling becomes inefficient \citep{DV2008, Schaye2015} and injecting
kinetic energy that adds velocity kicks to gas particles to remove them from
the inner regions of galactic discs \citep{Springel2003, Oppenheimer2006,
Vogelsberger2013}. These models have been quite successful in
reproducing the properties of galaxies in a broad sense
\citep{Vogelsberger2014, Vogelsberger2014N, Schaye2015, Springel2018, Marinacci2018, Naiman2018, Nelson2018, Pillepich2018}. However, they require
the fine tuning of quite a few number of free parameters that do not necessarily map to a set of physical processes. In many cases  they also require unrealistic values of SNe feedback
energy (${>10^{51}\,\text{erg}}$; \citealt{Guedes2011, Schaye2015}) or
excessively large gas outflow velocities \citep{Pillepich2018}.

 Only recently has there been a push to quantify the momentum input of SNe events by performing high resolution simulations that resolve the cooling radius and the Sedov-taylor phase of the explosion \citep{Kim2015, Martizzi2015, Walch2015, Haid2016}, which showed that the momentum amplification that can be achieved is of the order of $\sim 13-30$.  However, even with the momentum boost included, SNe feedback alone is not able to regulate star formation in galaxies \citep{Girichidis2015}, if the stars are assumed to form and explode in the high density peaks of the ISM. Only for random SNe positions is the energy injected in sufficiently low-density environments to reduce energy loses significantly. It enhances the effective coupling between the SNe feedback energy and the ISM leading to more realistic velocity dispersions and strong, high mass loaded winds leading to a global regulation of star formation. Similar results have been found by \citet{Kim2017}, who showed the importance of binary stars that leave their birth cloud and are able to deposit energy in lower density environments.  
 
In addition to SNe, young massive stars deposit large amounts
of energy in the form of  photons and stellar winds, which can have a significant dynamical impact on the ISM \citep{Leitherer1999, Murray2010,
Walch2012, Agertz2013}. \citet{Stinson2013} showed that the high energy photons emitted by
O and B stars can ionize and photoheat the surrounding regions, thereby preprocessing the sites of SNe explosions.  This helps regulate
star formation especially at high redshifts \citep{Kannan2014a}. This required full thermalization of the injected UV radiation energy close
to the source, which is not guaranteed as most of the energy of the photons is utilized to photoionize the gas. Radiation pressure, both direct UV and multiscattered infrared (IR), is
another mechanism invoked to drive significant outflows (${\sim
100\,\text{km s}^{-1}}$) \citep{Hopkins2011, Agertz2013, Hopkins2014}.
 It is however, unclear whether the the gas can actually trap the photons  efficiently. \citet{Krumholz2013} using a Flux Limited Diffusion (FLD) RT scheme, showed that as the gas accelerates in the presence of the gravitational potential of the disc, it becomes Rayleigh-Taylor unstable, creating channels through which photons escape, reducing the efficacy of this mechanism. Simulations performed with more accurate RT algorithms disagree with the previous calculations and show that it is indeed possible to drive outflows even when the gas becomes Rayleigh-Taylor unstable \citep{Davis2014, Zhang2017}. In any case, large optical depths (${\tau_\ts{IR}\sim 50}$) are required to effectively trap the photons
and boost the momentum injection to the levels required to efficiently suppress
star formation \citep{Roskar2014}.  These conditions are currently thought to only exist in galaxies with extremely high gas surface densities $\Sigma_{gas} \gtrsim 200 \ {\rm M}_\odot \ {\rm pc}^{-2}$. Alternatively, if enough radiation escapes
the star forming regions and the ISM of galaxies, it can in principle reduce
the gas cooling rates of the circum-galactic medium (CGM) thereby reducing gas inflows into the centers
of galaxies \citep{Cantalupo2010, Gnedin2012, Kannan2014b, Kannan2016a}.  

While
these works hint towards the importance of radiation fields, the crude nature of
these sub-grid models makes it difficult to gauge the exact
mechanisms and significance of radiation fields in regulating the star
formation rates of low mass galaxies. Therefore, full radiation hydrodynamic
simulations are necessary in order to gain a fundamental understanding of
stellar feedback \citep{Rosdahl2015b, Kim2017, Peters2017, Emerick2018}. \citet{Rosdahl2015b} using radiation hydrodynamic (RHD) isolated disc simulations showed that photoheating suppresses star formation by making
 the disc smooth and thick and by preventing of the formation of dense clouds (rather than dispersing them). Radiation pressure (both UV and multi-scattering IR) was shown to be
 unimportant. The need to simulate the entire disc necessitated relatively low resolutions  ($\sim 20-30 \ \rm{pc}$), which meant that both the Str\"omgren radius around young massive stars and the Sedov-Taylor phase of the SNe explosion were only partially resolved. \citet{Peters2017} showed that radiation fields in combination with stellar winds can regulate star formation in solar-neighbourhood like environments. However, the effect of radiation pressure was left unexplored.   
 
 In this paper, we perform high resolution ($\Delta x \sim 0.45$ pc ; $M_{\rm gas} = 10 \ {\rm M}_\odot$) simulations of a patch of the ISM to investigate and quantify the role of various stellar feedback processes like SNe, photoheating and radiation pressure (both UV and Multi-scattered IR) in low gas surface density galaxies. These high resolutions ensure that the relevant spatial and mass scales are sufficiently resolved, thereby providing an accurate picture of stellar feedback.  The paper is structured as follows. In Section \ref{sec:methods} we outline the simulations performed. Section \ref{sec:results} describes the results obtained. The interpretation of the results is presented in Section~\ref{sec:discussion} and finally, we present our conclusions in Section \ref{sec:conclusions}.

\section{Methods}
\label{sec:methods}
Our simulation setup is the same as the one described in \citet{Simpson2016} which in turn in based on the setup described in \citet{Creasey2013}. Briefly it consists of a column of stratified gas intended to represent a small portion of a galactic disk. The domain dimensions are $1 \times 1 \times 10 \ {\rm kpc}$ and we impose periodic boundaries along the two short ($x$ $\&$ $y$) axes and outflow boundaries along the long ($z$) axis. The gravitational forces are computed as a sum of self-gravity and an external potential mimicking the pre-existing stellar density field. The self-gravity is computed using a tree based algorithm. An  adaptive softening  is  used  for
gas  cells  with  a  minimum  value  of $\epsilon_{\rm gas} = 0.165 \ \rm{pc}$. The collisionless stellar particles have a fixed softening of $\epsilon_{\rm star} = 0.165 \ {\rm pc}$. The stellar density field is proportional to the initial gas density ($\rho_{{\rm gas},0}$) field via the assumed gas fraction ($f_g$); $\rho_\star = \rho_{{\rm gas},0}(f_g^{-1}-1)$.  The initial gas density profile is given by
\begin{equation}
\rho_{{\rm gas},0} = \frac{\Sigma_0}{2b_0} \sech^2\left( \frac{h}{b_0} \right) \ ,
\end{equation}
where the gas surface density is set to $\Sigma_0 = 10 \ {\rm M}_\odot \ {\rm pc}^{-2}$, the scale height is $b_0 = 100 \ {\rm pc}$ and the gas fraction is $f_g=0.1$. Note that the gas fraction is smaller than the fiducial value of the solar neighbourhood ($f_g \simeq 0.25$),  and the inverse correlation between the stellar potential and the gas fraction means that our gravitational potential is a factor of $2.5$ larger than solar neighbourhood conditions. This has important implications for the star formation rates obtained in our simulations (see Section~\ref{sec:results} for more details). We impose a minimum density threshold of $10^{-20} \ {\rm M}_\odot \ {\rm pc}^{-3}$. 

The simulations are performed with  {\sc Arepo-RT} \citep{Kannan2018} a radiation hydrodynamic extension to the moving mesh code {\sc Arepo} \citep{Springel2010, Pakmor2016}.  {\sc Arepo} provides a quasi-lagrangian solution to the hydrodynamic equations by solving them at interfaces between moving mesh cells in the rest frame of the interface. We assume a thermal adiabatic index of $\gamma = 5/3$. A minimum allowed temperature of 5 K is adopted. The initial setup consists of $10^6$ gas cells,
concentrated in the mid-plane, but also comprising a Cartesian background mesh with a cell length of $43.5$ pc up to $1$ kpc and of $90.9$ pc beyond. Refinement and derefinement of the mesh is applied to maintain roughly constant cell masses to within a factor of two of the target gas mass of $10 \ {\rm M}_\odot$, subject to the constraints that cell volumes are approximately limited between $0.1 \ {\rm pc}^3$ and $7.19 \times10^5 \ {\rm pc}^3$, a maximum volume ratio of 8 between adjacent cells is maintained, and cell diameters are required to be no larger than $1/4$ of the Jeans length. To keep the
Jeans length resolved after gravitational collapse has reached the minimum allowed cell volume,  an effective pressure
floor in the Riemann
solver equal to $4^2$
times the Jeans pressure is imposed \citep{Machacek2001}.

The RHD module solves the moment-based radiative transfer equations using the M1 closure relation. We achieve second order accuracy by using a slope-limited linear spatial extrapolation and a first order time prediction step to obtain the values of the primitive variables on both sides of the cell interface, which are then used to solve the Riemann problem at the interface. This allows the code to be extremely accurate and have excellent diffusivity control.  The M1 closure method is fully local in nature, such that the computational cost is independent of the number of sources and only depends on the number of resolution elements in the simulation.

We use the chemistry and cooling network outlined in \citet{Smith2014} . It solves the hydrogen chemistry, including ${\rm H}_2$ \citep{Glover2007a, Glover2007b} and has a simple treatment for CO chemistry \citep{Nelson1997, Glover2012}. The carbon , oxygen and helium abundances are the same as used in \citet{Smith2014}. The dust to gas ratio is assumed to be $f_d = 0.01$. We do not use any external radiation field as they will be generated self-consistently in our simulations. Metal cooling of high-temperature gas assuming collisional ionization equilibrium is also included \citep{Gnat2012, Walch2015} assuming a constant solar gas metallicity.

The chemistry network is coupled to radiation fields using a multi-frequency approach. We use six frequency bins: the IR band ($0.1-1 \ {\rm eV}$), optical band ($1-11.2 \ {\rm eV}$), the Lyman-Werner (LW; ${\rm H}_2$ dissociation) band ($11.2-13.6\ {\rm eV}$), hydrogen ionization band   ($13.6-15.2\ {\rm eV}$), \ion{H}{} and \ion{H}{$_2$} ionization band  ($15.2-24.6\ {\rm eV}$) and finally the \ion{He}{} ionization band ($24.6-100.0\ {\rm eV}$). The dust opacity to IR radiation is set to $\kappa_{\rm IR} = 10 \ {\rm cm}^2 {\rm g}^{-1}$ and the opacity for all other radiation bins is  $\kappa_{\rm UV} = 1000 \ {\rm cm}^2 {\rm g}^{-1}$.  The photoionization and photoheating rates for each bin are calculated as described in Section 3.2.1 of \citet{Kannan2018}. 

Accurately simulating ${\rm H}_2$ thermochemistry is quite tricky as only about $10 \%$ of the LW absorption leads to dissociation and the rest of the photons are destroyed without contributing to photodissociation \citep{Draine1996, Sternberg2014}. The absorption rate is highly dependent on the wavelength of the LW band \citep{Haiman2000} because more resonant bands become optically thick at high ${\rm H}_2$ column densities, and dissociation is quashed, while bands with weaker absorption can still penetrate the cloud. Hence, ${\rm H}_2$ self-shielding functions calculated from experiments are given in terms of the column density of ${\rm H}_2$ \citep{Gnedin2014}. Unfortunately, our formulation of the RT equations does not track the optical depth of individual rays. Therefore, we resort to using the method described in \citet{Nickerson2018}  and boost the destruction of LW photons by a constant factor to incorporate the fact that only a fraction of LW photon absorption leads to ${\rm H}_2$ dissociation. As the LW photons propagate through gas cells, their repeated destruction mimics the column density variation of ${\rm H}_2$ destruction rates. A test for this scheme is presented in Appendix \ref{sec:h2}. 

Stars are formed following a standard stochastic approach. The density threshold for star formation is set to $n = 100 \ {\rm cm}^{-3}$. The star formation rate of a cell `i' above this threshold is set to
\begin{equation}
{\rm sfr}_i = \epsilon_{\rm ff} \frac{m_i}{t_{\rm ff}} \ ,
\end{equation}
where $\epsilon_{\rm ff}$ is the star formation efficiency per free fall time of gas (set to $0.02$) and $t_{\rm ff}$ is the free fall time of the gas defined as
\begin{equation}
t_{\rm ff} = \sqrt{\frac{3 \pi}{32 G \rho_i}} \ .
\end{equation}
The probability of a cell forming a star is then given by 
\begin{equation}
p_i = {\rm sfr}_i \frac{\Delta t}{m_\star}  \ , \  {\rm where}  \ m_\star = {\rm min} \{ m_i, m_{\rm max} \} \ . 
\end{equation}
Collisionless particles of mass $m_\star$ , representing stellar populations, are formed stochastically from the gas, with the probability of forming one drawn from a Poisson distribution.  Note that if $m_i = m_\star$, then the whole cell is converted to stars else part of the cell mass is converted into stars with the the maximum stellar mass set to $m_{\rm max} = 20 \ {\rm M}_\odot$. 

 We assume a Chabrier \citep{Chabrier2003} stellar initial mass function (IMF). Stars with initial mass ${\rm M}_\star \geq 8 \ {\rm M}_\odot$ are assumed to explode as SNe at the end of their lifetime. This sets the SNe rate  ${\rm SNR} \sim 1$ per $100 \ {\rm M}_\odot$ of stars formed. The extremely high mass resolution of our simulations necessitates the need for a probabilistic approach to stellar feedback. As soon as a star is formed we tag that stellar particle to go SNe in a probabilistic manner. The probability of a star going SNe is given as
 \begin{equation}
 p_{\rm SNe} = {\rm SNR} \frac{m_\star}{100 \ {\rm M}_\odot} \ .
 \end{equation}
 We then enforce that only the tagged SNe particles emit radiation fields. This ensures that irrespective of the mass of the stellar particle, it emits radiation equivalent to $100 \ {\rm M}_\odot$ of new stars formed.  The radiation luminosity and spectra are obtained from \citet{Bruzual2003}. The time delay between star formation and the SNe event is set to $5 \ {\rm Myr}$. We assume that the stars only emit radiation during this time and as soon as they go SNe their radiation output stops. This is a good approximation as stellar population synthesis models predict a precipitous drop in the radiation output after about $3 \ \rm{Myr}$, when the most massive stars start to die off.  SNe are modeled as discrete explosions of $10^{51} \ {\rm erg}$  deposited as purely thermal energy into the 32 closest cells to the explosion position. Explosion events are only added to the mesh when all gas cells are synchronized; the maximum allowed timestep is $0.1 \  {\rm Myr}$.  The high mass and spatial resolution of our simulations ensures that the right momentum boost is recovered \citep{Simpson2015, Simpson2016}.

\begin{figure*} 
\includegraphics[scale=0.4]{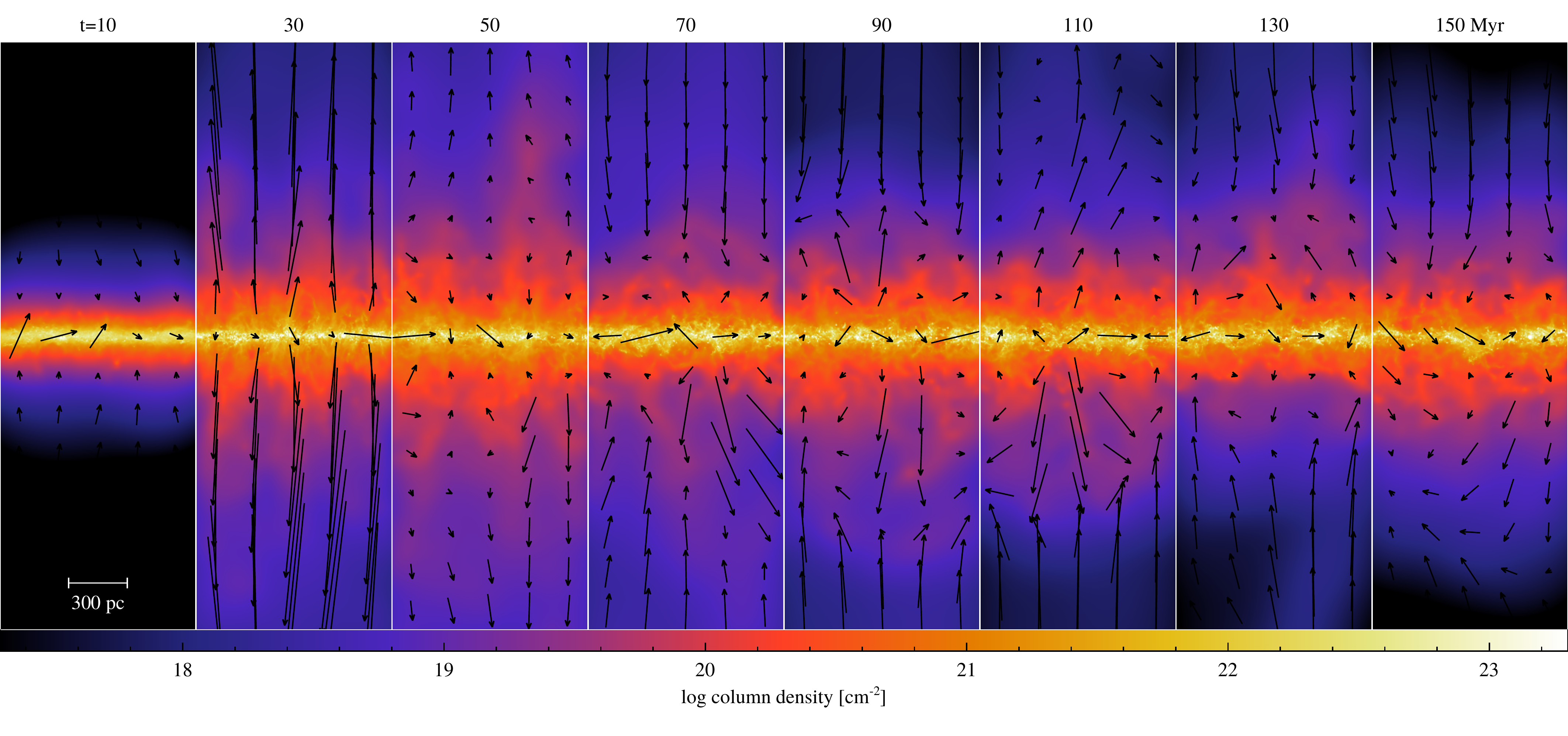}  
\caption{Map of the column density along the line of sight in the SN simulation at $t=10$ (first column), $30$ (second column), $50$ (third column), $70$ (fourth column), $90$ (fifth column), $110$ (sixth column), $130$ (seventh column) and $150$ (eighth column) Myr. The dimensions of the box shown in the image are $1 \times 3$ kpc. The black arrows indicate the direction of the velocity field with the length of the arrows indicating the magnitude of the velocity field. The initial starbusrt drives large scale ($\sim 3-4$ kpc) outflows ($t=30-50$ Myr), followed by a period of infall ($t=50-70$ Myr), after which the disc settles down with a small scale fountain flow ($ \lesssim \ \rm{kpc} $) operating from $t=70-150$ Myr.}
\label{fig:densmapfd}    
\end{figure*}

\begin{figure*} 
\includegraphics[scale=0.4]{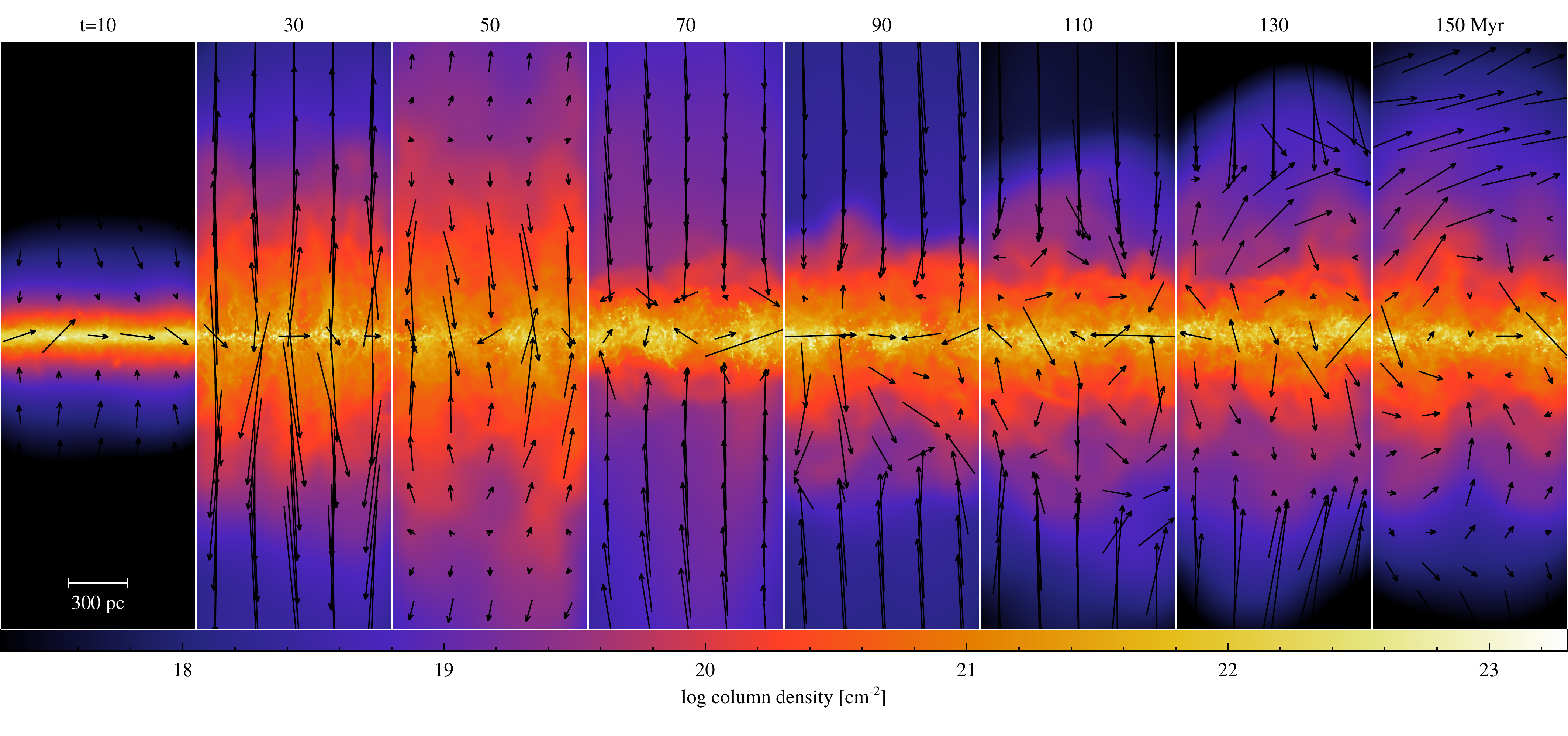} 
\caption{Same as Fig. \ref{fig:densmapfd} but for the PH simulation.}
\label{fig:densmapph}  
\end{figure*}

\begin{figure*}

 \includegraphics[scale=0.4]{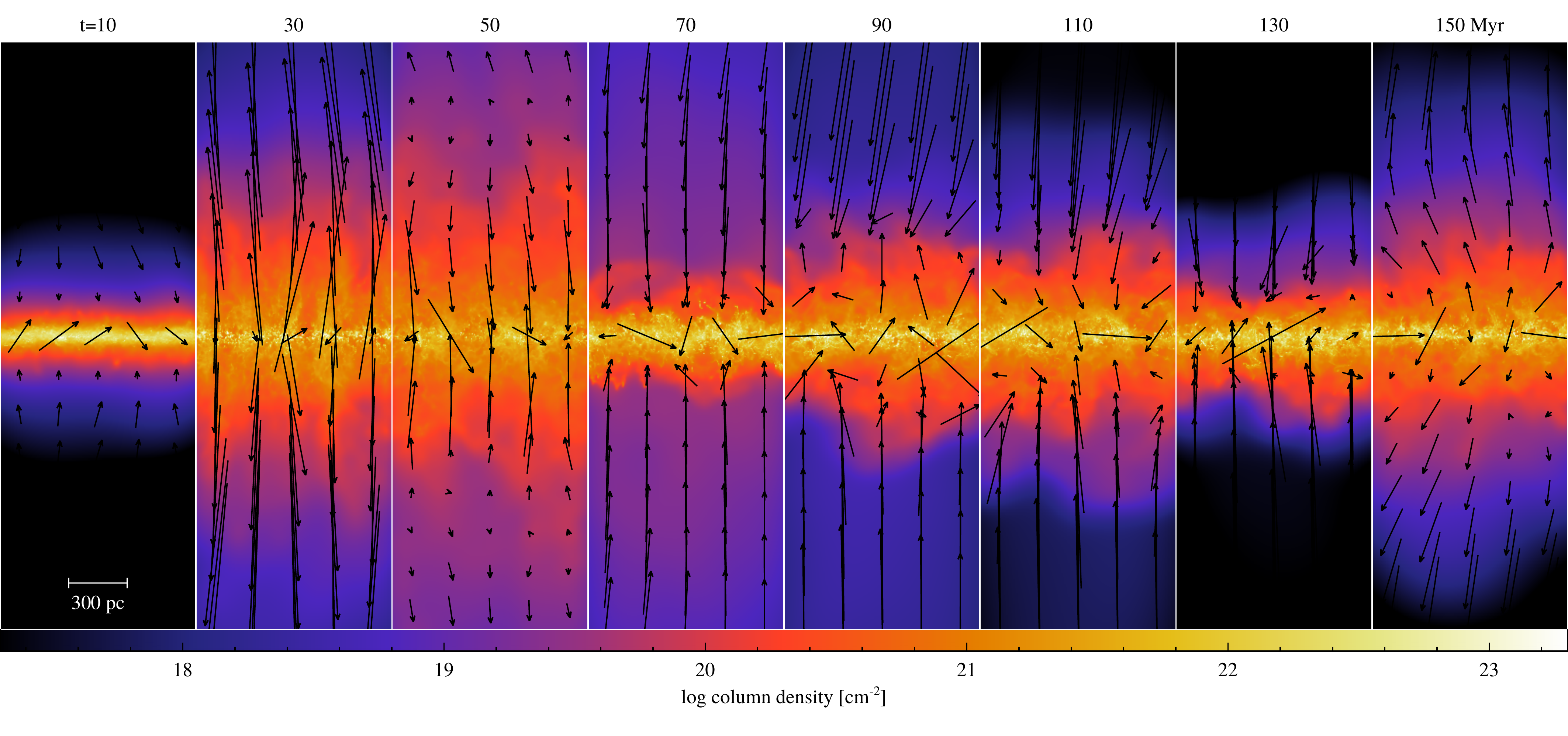}
 \caption{Same as Fig. \ref{fig:densmapfd} but for the RP simulation.}
 \label{fig:densmaprp}
\end{figure*}

\begin{figure*}
 \includegraphics[scale=0.4]{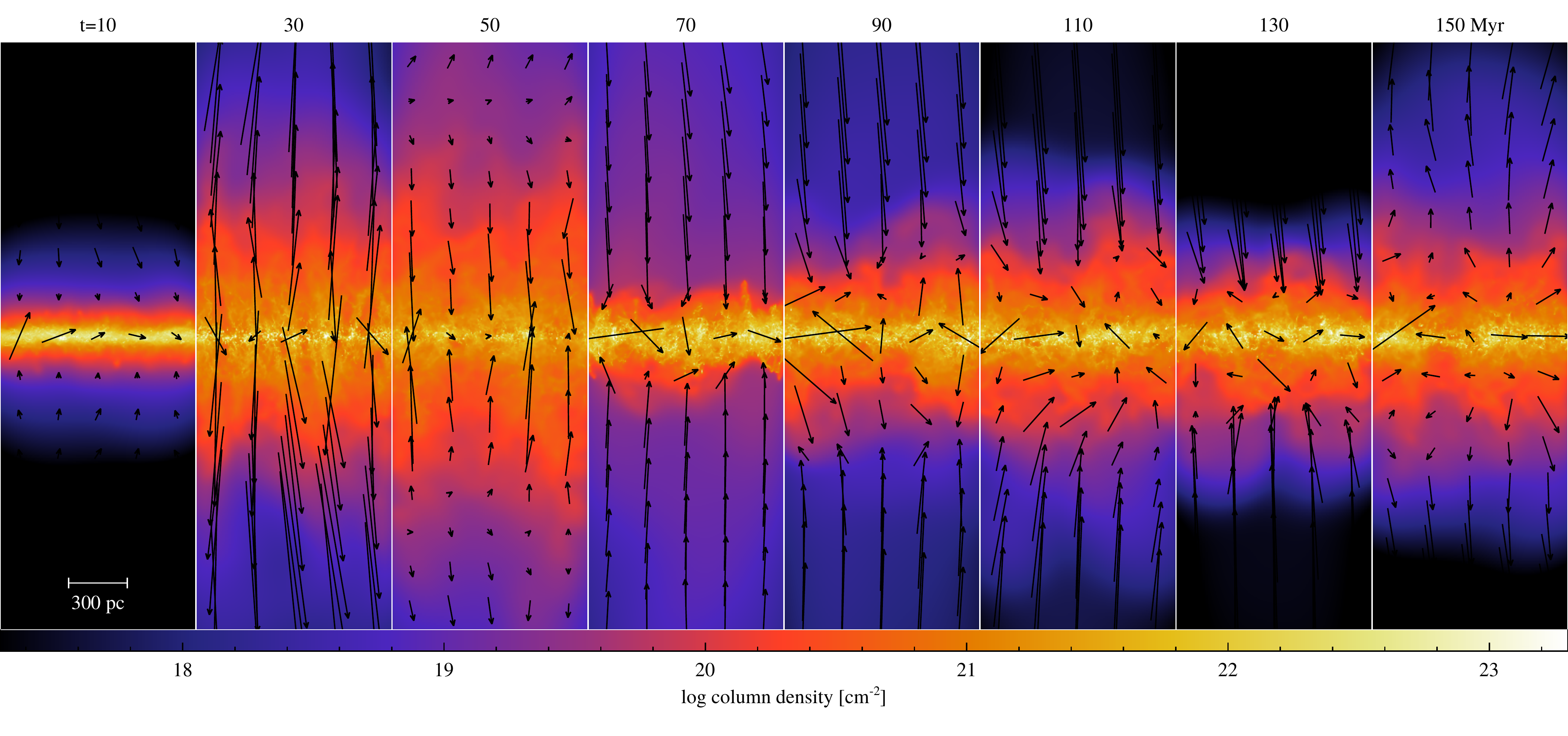}
 \caption{Same as Fig. \ref{fig:densmapfd} but for the IR simulation.}
 \label{fig:densmapir}
\end{figure*}

 The amount of UV ionizing photons per $100 \ {\rm M}_\odot$ of stars formed is about $\dot{N}_\gamma = 5 \times 10^{48} \ {\rm photons \ s}^{-1}$ \citep{Bruzual2003}. The Str\"omgren radius ($r_s$), assuming full ionization within $r_s$, is given by
 \begin{equation}
 r_s = \left(\frac{3\dot{N}_\gamma}{4 \pi \alpha_B n_{\rm H}^2}\right)^{1/3} \ ,
 \end{equation}
 where $\alpha_B = 2.63 \times 10^{-13} \ {\rm cm}^3 \ {\rm s}^{-1}$ is the Case B recombination rate for hydrogen atoms at $T=10^4 \ {\rm K}$. The minimum cell sizes in our simulation reaches about $0.1 \ {\rm pc}^3$ which equates to maximum  density at which the Str\"omgren radius is resolvable to $n = 2.6 \times 10^3 \ {\rm cm}^{-3}$.  Above this density $r_s$  will be unresolved and the effect of photoheating will be underestimated. To increase the probability of resolving the $r_s$, we inject all the photons in the cell closest to the star particle. Additionally, the direction of the photon flux (${\bf F}_r$) is set to be radially outward from the star particle and the magnitude is $|{\bf F}_r| = \tilde{c} E_r$, where $E_r$ is the photon energy density and $\tilde{c}$ is the reduced speed of light, which in our simulations is set to $10^3 \ {\rm km \ s}^{-1}$.  This overcomes the issues mentioned in \citet{Hopkins2018z} by ensuring that the full radiation pressure force is accounted for even if the cell optical depth is larger than one. However, this leads to anisotropic pressure forces around a star particle, but this is mitigated by the fact that we form a large number of stars during the simulation and each random orientation adds up to create an isotropic pressure force. 
 
 We perform four different simulations; SN: only the SNe feedback is active; PH: SN + photoheating from UV sources is active; RP: PH + radiation pressure from just single scattering UV radiation is active and finally IR: RP + effect of multi scattered IR radiation. Each of these simulations are run for $t=150 \ {\rm Myr}$. In this paper we have decided to focus only on the effect of radiation fields and hence our simulations neglect the effects
 of other important ingredients such as the magnetic field, winds from massive stars, and cosmic rays. 
\section{Results}
\label{sec:results}

\begin{figure} 
\includegraphics[width=\columnwidth]{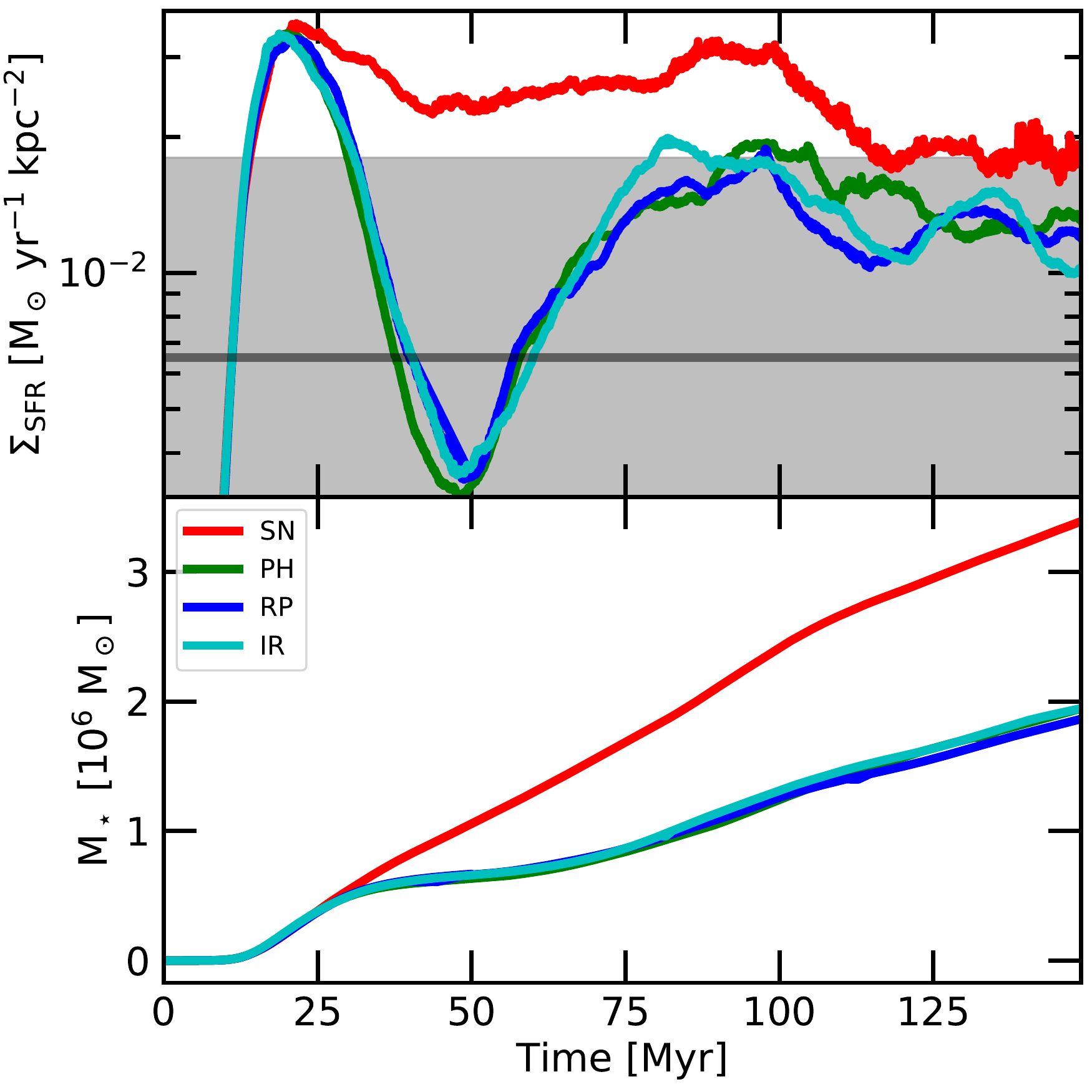} 
\caption{The star formation rate surface density (top panel) and the total amount of stellar mass faormed (bottom panel) as a fucntion of time (in Myr) in the SN (red curves), PH(green curves), RP (blue curves) and IR (cyan curves) simulations. The solid grey line depicts the expected star formation rate from the Kennicutt-Schmidt relation \citep{Kennicutt2012} and the grey shaded region is the factor of three observational error in the observed relation. After the initial starbusrt, the simulations with ESF show a steep drop in the SFR, which slowly bounces back and regulates to a value by a factor of $\sim 2$ lower than in SN run.The total amount of stars formed is also reduced by a factor of $\sim 1.5$ in the runs with ESF. The negligible difference between the PH, RP and IR runs shows the inability of radiation pressure (both single and multi-scattering) to regulate star formation in low surface density galaxies.}
\label{fig:sfr}
\end{figure}

We begin by looking at the morphological evolution of the disc in our simulations. Fig.~\ref{fig:densmapfd} shows the  $x-z$ ($1 \times 3 \ {\rm kpc}$) map of the column density integrated along the $y-$ axis in the SN simulation at $t=10$ (first column), $30$ (second column), $50$ (third column), $70$ (fourth column), $90$ (fifth column), $110$ (sixth column), $130$ (seventh column) and $150$ (eighth column) Myr.  The black arrows indicate the direction of the velocity field with the length of the arrows representing is magnitude.  During the initial $10 \ {\rm Myr}$, the disk cools and contracts before the first stars form and explode as SNe. This causes a small scale inflow which reduces the scale height of the disk and induces a starburst.  As the stars begin injecting feedback energy into the ISM, it starts to drive an outflow.  This outflow period lasts up to $\sim 50 \ {\rm Myr}$, during which time the disc scale height increases. The outflow reaches up to $\sim 4 \ {\rm kpc}$,  stall and begins to fall back to the disc. This inflow period lasts up to $75 \ {\rm Myr}$. The disk then settles down and forms stars at a constant rate which creates a small scale ($\lesssim 1 \ {\rm kpc}$) fountain flow. The outflow velocities generated during the star-burst phase of evolution is generally higher compared to the fountain flow phase.

The morphological evolution of the disc in the PH (Fig.~\ref{fig:densmapph}), RP (Fig.~\ref{fig:densmaprp}) and IR (Fig.~\ref{fig:densmapir}) simulations is qualitatively quite similar, with a prominent starburst driven outflow phase followed by an inflow phase and a small-scale fountain flow phase. However, there are some interesting differences between the SN run and the runs with radiation fields (or runs with early stellar feedback or ESF runs). Throughout the simulation, the central high density disc remains relatively unperturbed in the SN run  while the ESF runs manage to make the ISM more clustered creating low density holes and filamentary channels through which material can be ejected without hindrance. This allows the ESF runs to drive gas flows to larger heights above the disc. This difference is clearly visible during the initial starburst period ($t = 30-50 \ {\rm Myr}$). They are also able to entrain more high density material in the outflow in both the starburst and fountain flow period of the evolution. These column density maps shows that that including radiation fields makes stellar feedback qualitatively more effective. Most of the difference is seen between the SN and PH runs and there is very little difference between the PH, RP and IR runs, implying that  photoheating is the primary mechanism through which radiation fields interact with the ISM and radiation pressure has minimal effect in regulating star formation and driving outflows in our model.

\begin{figure} 
\includegraphics[width=\columnwidth]{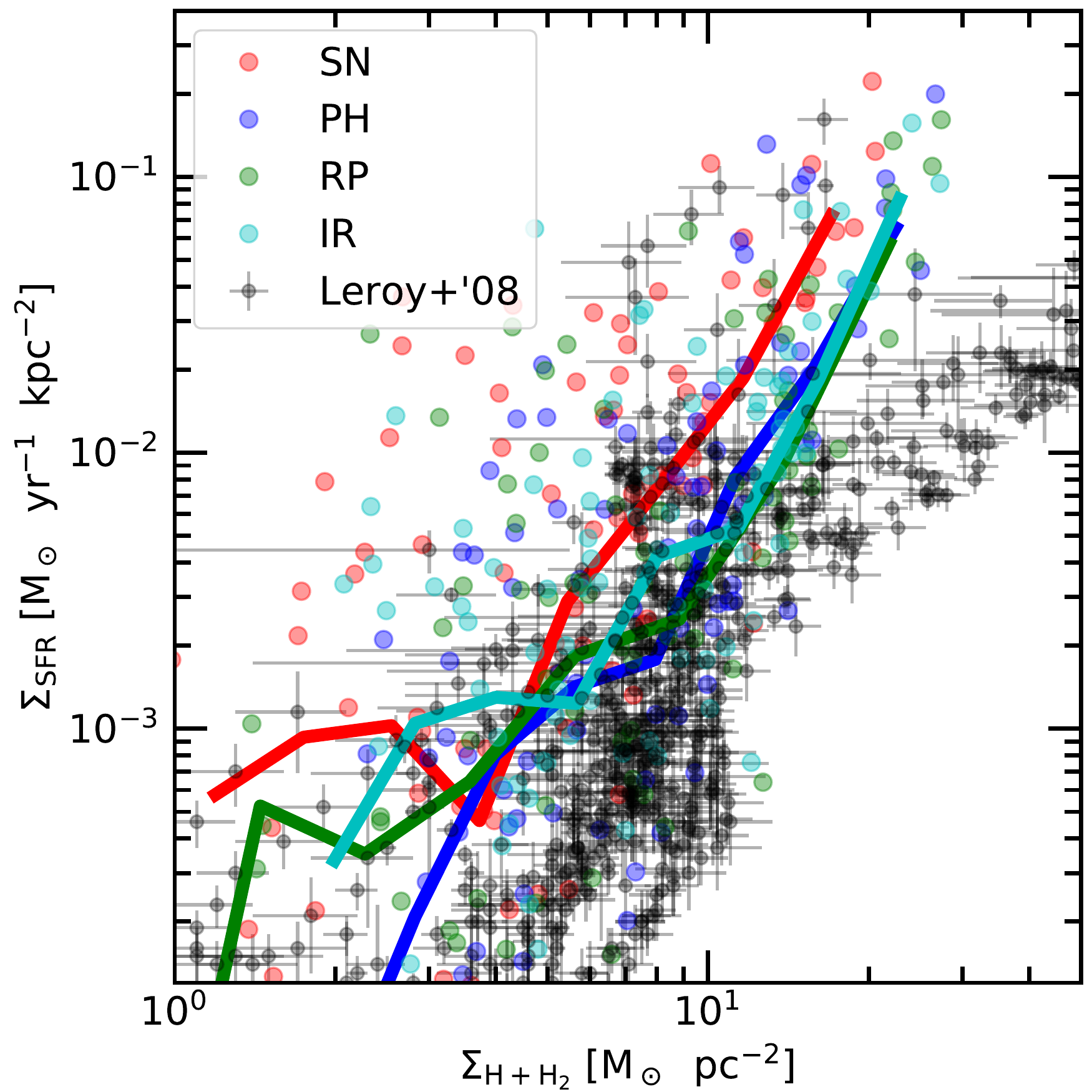}   
\caption{The star formation rate surface density as a function of the hydrogen (H+H$_2$) surface density in the SN (red), PH (green), RP (blue) and IR (cyan) simulations. The data is taken at the simulation time of $t=150$ Myr. The individual points are obtained by dividing the disc into patches of size $100\times100$ pc and calculating both quantities for each patch. The solid lines denote the median for each simulation. For comparison, the observational estimates from \citet{Leroy2008} are plotted with black circles. The runs with ESF tend to bring the values close to the observed estimates.} 
\label{fig:KS}  
\end{figure}

\begin{figure*} 
\includegraphics[scale=0.39]{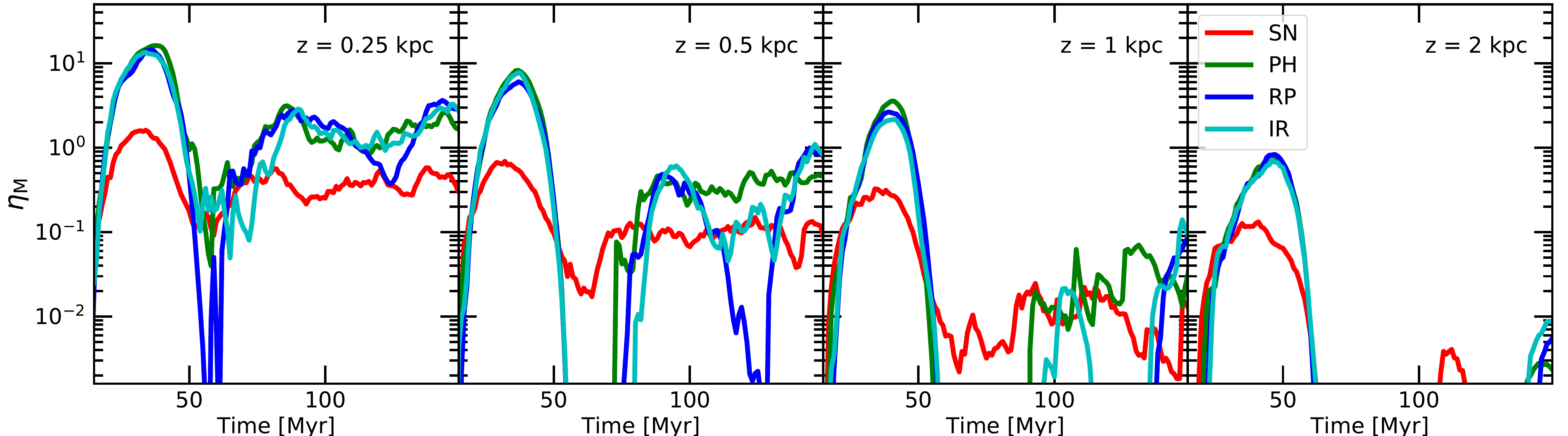}
\caption{The mass loading factor (defined as $\eta_{\rm M}  = \dot{M}_{\rm out}/\dot{M}_\star$) as a function of time at z$=0.25$ (first panel), z$=0.5$ (second panel), z$=1.0$ (third panel) and z$=2.0$ (fourth panel) kpc above the disc for the SN (red curves), PH (green curves), RP (blue curves) and IR (cyan curves) simulations. $\eta_{\rm M}$ generally decreases with z for all the runs and the runs with ESF generally have higher  $\eta_{\rm M}$'s at all times.}
\label{fig:mload}
\end{figure*}

\begin{figure*}   
\includegraphics[scale=0.39]{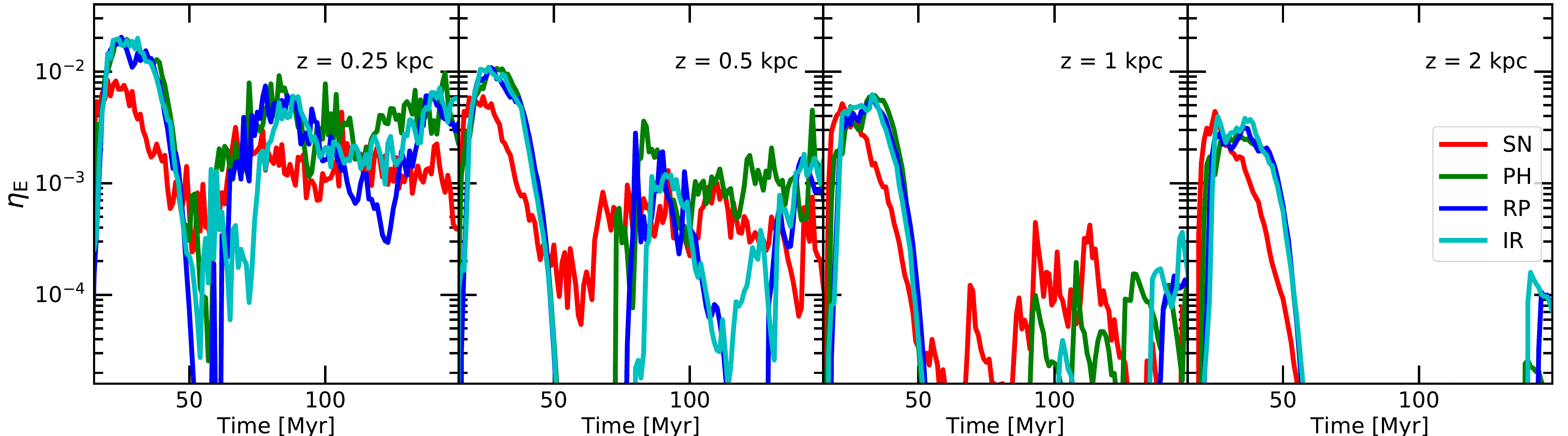}
\caption{The energy loading factor (defined as $\eta_{\rm E}  = \dot{E}_{\rm out}/\dot{E}_{\rm SNe}$) as a function of time at z$=0.25$ (first panel), z$=0.5$ (second panel), z$=1.0$ (third panel) and z$=2.0$ (fourth panel) kpc above the disc for the SN (red curves), PH (green curves), RP (blue curves) and IR (cyan curves) simulations.}
\label{fig:Eload}
\end{figure*}  

We begin a more quantitative comparison by plotting the star formation rate surface density (top panel) and the total stellar mass formed (bottom panel) with time in the SN (red curves), PH (green curves), RP (blue curves) and IR (cyan curves) simulations (Fig.~\ref{fig:sfr}). The solid grey line depicts the expected star formation rate from the Kennicutt-Schmidt relation \citep{Kennicutt1998, Kennicutt2012} and the grey shaded region is the factor of three observational error in the observed relation. The amplitude of the initial starburst is similar in all simulations. This is expected, because the stars have not had the time to feed energy back into the ISM. After about $20 \ {\rm Myr}$, the SFR curves start to deviate. The SN run continues to form stars at a very high rate ($\sim 0.02 - 0.03 \  {\rm M}_\odot \ {\rm yr}^{-1} \ {\rm kpc}^{-2}$), while all three runs with radiation fields show a precipitous drop in the SFR's, with the minimum as low as  $\sim 0.003 \  {\rm M}_\odot \ {\rm yr}^{-1} \ {\rm kpc}^{-2}$.  By about $75 \ {\rm Myr}$ the SFR's bounce back and saturate to a value that is about $1.5 - 2$ lower than the SFRs obtained from the SN simulation. The SFR in the SN run always remains above the observed value throughout the simulation, matching the SFR estimates from the simulation including only SNe feedback presented in \citet{Peters2017}. We note that the drop off in the SFR surface density at about $\sim 100  \ {\rm Myr}$ is due to the significant decrease in the gas surface density by $\sim 30 \%$ due to the conversion of gas into stars. The ESF runs on the other hand manage to contain the SF within observational limits, even with higher gas surface density at late times, though it must be noted that it is at the upper end of the observed error margin. The total stellar mass formed in the ESF runs are also about factor of two lower than the SN simulation.  There is very little difference between the PH, RP and IR runs, implying that radiation pressure, both UV and IR, is unimportant in these low density environments.  Photoheating, therefore, emerges as the most important early stellar feedback mechanism, which is in agreement with previous works \citep{Stinson2013, Kannan2014a, Sales2014, Rosdahl2015b}.

Fig.~\ref{fig:KS} shows the star formation rate surface density as a function of the gas (\ion{H}{} + \ion{H}{$_2$}) surface density at $t = 150 \ {\rm Myr}$, for the SN (red curves), PH (green curves), RP (blue curves) and IR (cyan curves) simulations. The observational data points (black circles) are obtained from observations of  $23$ nearby galaxies as outlined in \citet{Leroy2008}.   The individual data points from our simulations are obtained from dividing the disk into $100 \times 100 \ {\rm pc}$ chunks and calculating both the quantities for each chunk.  This allows us to study the KS relation in a wide variety of environments. The solid lines indicate the median values, which generally lie within the cloud of observed data points. Including radiation fields brings the simulated KS relation closer to the observed values. We do however, somewhat overshoot the relation possibly hinting at missing important physical processes like stellar winds, cosmic rays or the effect of binary stars leaving their birth clouds and exploding in low density environments.

   \begin{figure*}  
\includegraphics[scale=0.45]{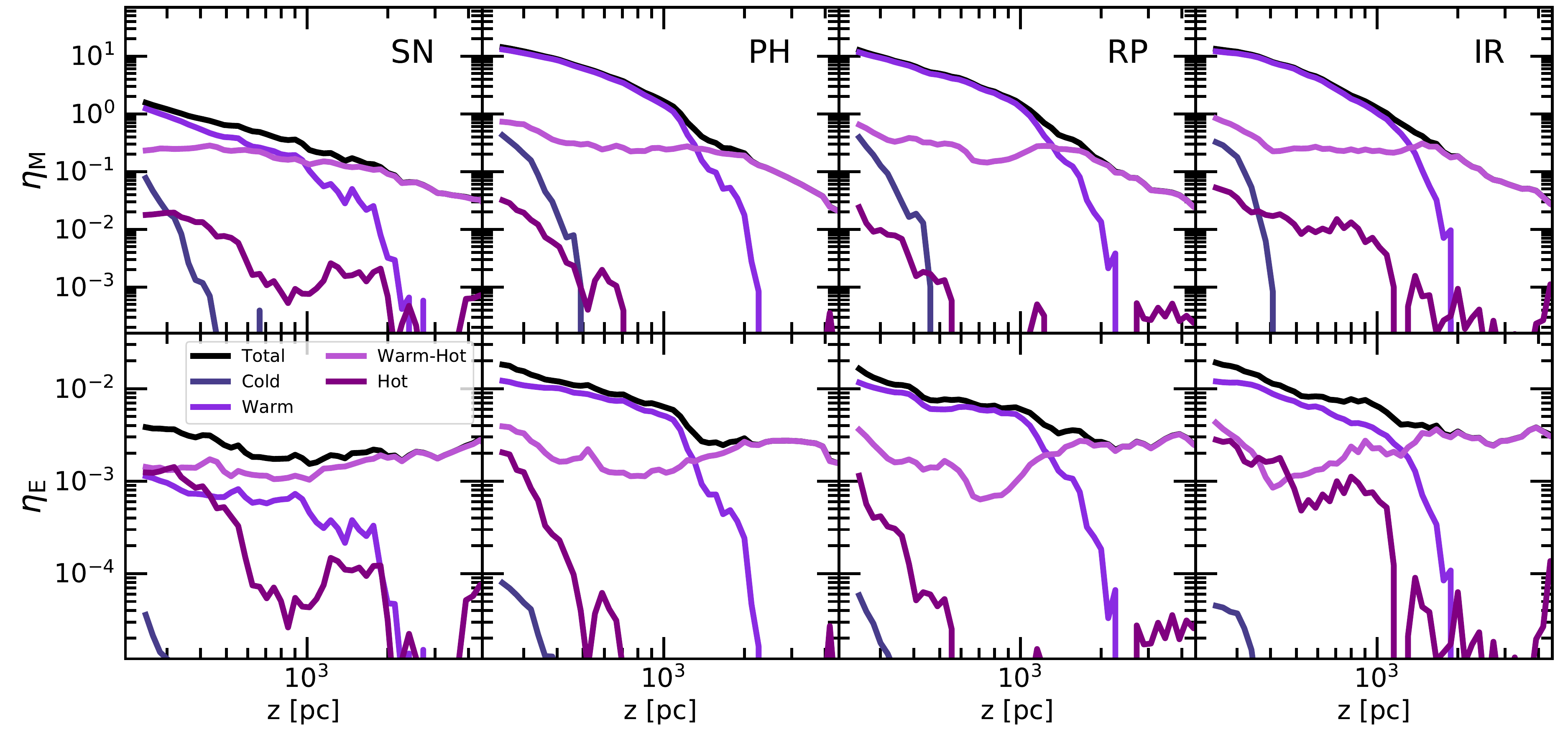}
\caption{The mass (top panels) and energy loading (bottom panels) factors as a function of height from the disc (z) for cold (${\rm T} \leq 300 \ {\rm K}$), warm ($300 < {\rm T} \leq 8000 \ {\rm K}$), warm-hot ($8000 < {\rm T} \leq 3\times 10^5 \ {\rm K}$) and hot gas (${\rm T} > 3\times 10^5 \ {\rm K}$) during the outflow phase of the simulation, $t=30$ Myr. The first column shows the values for the SN simulation, second column; PH, third column; RP and fourth column; IR.  Most of the mass and energy of the outflow is in the warm and warm-hot phases, with the cold and hot gas being subdominant. The increase in mass and energy loading factors in the runs with ESF is primarily driven by the increase in outflow of the warm phase.}
\label{fig:dcT30}  
\end{figure*}

\begin{figure*} 
\includegraphics[scale=0.45]{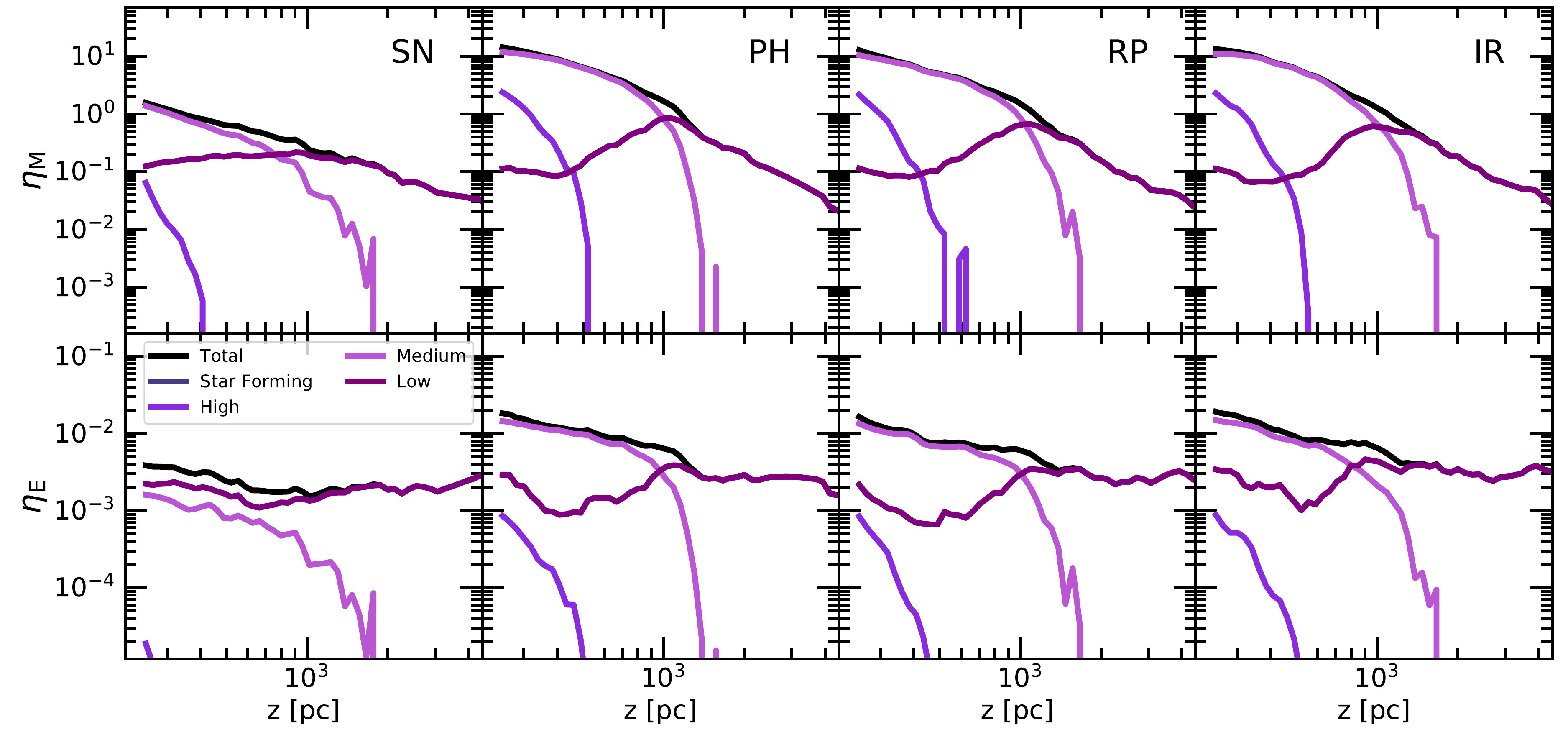}
\caption{The mass (top panels) and energy loading (bottom panels) factors as a function of height from the disc (z) for star-forming ($n \geq 10^2 \ {\rm cm}^{-3}$), high ($1\leq n< 10^2  \ {\rm cm}^{-3}$), medium ($10^{-2} \leq n< 1  \ {\rm cm}^{-3}$) and low ($n<10^{-2} \ {\rm cm}^{-3}$) density gas during the outflow flow phase of the simulation, $t=30$ Myr. The first column shows the values for the SN simulation, second column; PH, third column; RP and fourth column IR.}
\label{fig:dcn30}  
\end{figure*}

 We now turn our attention to the key quantities that describe a SNe driven wind, the mass ($\eta_{\rm M}$) and energy ($\eta_{\rm E}$) loading factors. The mass loading factor at a height $z$ from the plane of the disc is defined as the ratio between the outgoing mass flux and the star formation rate:
 \begin{equation}
  \eta_{\rm M}|_z = \frac{\dot{\rm M}_{\rm out}(z)}{\dot{\rm M}_\star} \ .
 \end{equation}
Similarly, the energy loading factor at $z$ is defined as the ratio of the total energy flux carried away by the wind to the energy injection rate by SNe:
 \begin{equation}
  \eta_{\rm E}|_z = \frac{\dot{E}_{\rm kin,out} + \dot{E}_{\rm therm,out}}{E_{\rm SN}\dot{\rm M}_\star} \ ,
 \end{equation}
 where $E_{SN} = 10^{51} \ {\rm erg} / 100 \ {\rm M}_\odot$. ${\rm M}_{\rm out}$ and ${\rm E}_{\rm out}$ are calculated only for gas that has a $z$ velocity vector pointing away from the disc midplane (i.e., we do include any inflowing gas in our analysis of mass and energy loading factors). It is important to quantify these quantities as sub-grid feedback models generally prescribe the mass, energy and metal loading factors and tune them in order to reproduce the observed galaxy properties \citep{Vogelsberger2013}. 
 
 Fig.~\ref{fig:mload} shows the mass loading factor at $0.25  \ {\rm kpc}$ (first panel), $0.5  \ {\rm kpc}$ (second panel), $1  \ {\rm kpc}$ (third panel) and $2  \ {\rm kpc}$ (fourth panel) above the disc for the SN (red curves), PH (green curves), RP (green curves) and IR (cyan curves) simulations. The behaviour of $\eta_{\rm M}$ is qualitatively similar in all the runs, with its value peaking during the initial starburst-driven outflow phase, followed by a decrease during the inflow phase and a rebound, at least close to the disc, during the fountain flow period. During the initial starburst period, the SN simulation drives outflows with a mass loading of $\sim 2$ at $z=0.25 \ {\rm kpc}$ which gradually drops to a value of about $\sim 0.1$ at $z=2 \ {\rm kpc}$. This weak outflow leaves the disc relatively unaltered, which allows the SF to continue at a relatively high rate even during the post-starburst period. The slight decrease in the SFR leads to a nominal decrease in $\eta_{\rm M}$ at low $z$ ($\lesssim 1 \ {\rm kpc} $). The mass loading at high $z$'s, however, shows a large drop off because the lower SFR's combined with a puffier disc makes SNe feedback less efficient. Under these conditions feedback is able to drive only a small-scale fountain flow up to a height of $z\sim 1 \ {\rm kpc}$.

\begin{figure*}   
\includegraphics[scale=0.5]{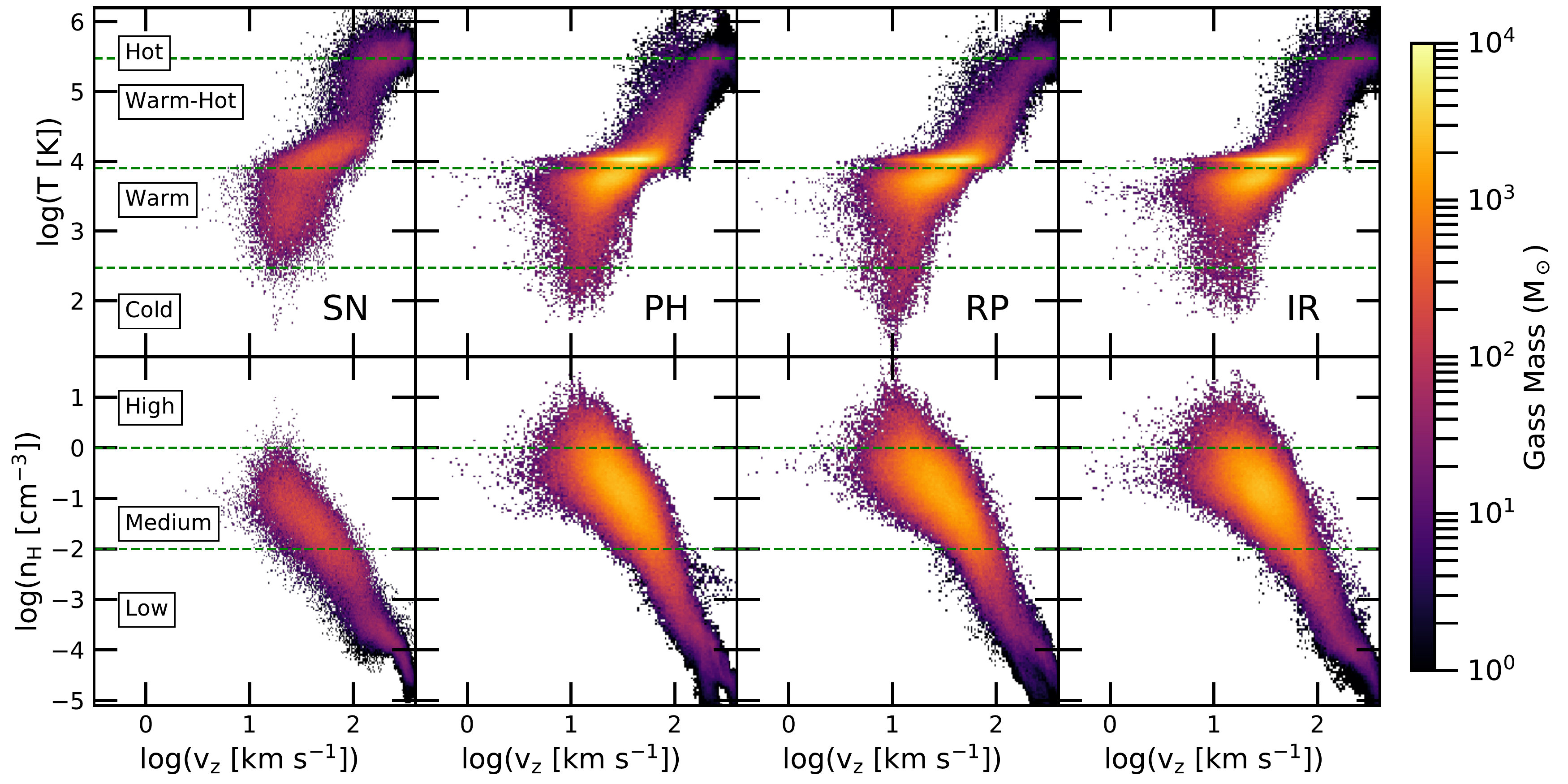}  
\caption{The temperature (top panels) and density (bottom panels) histograms of the outflowing gas as a function of the outflow velocity at $t=30$ Myr for the SN (first column), PH (second column), RP (third column) and IR (fourth column) simulations. }
\label{fig:vel30}  
\end{figure*}

The quantitative picture in the ESF runs is quite different. The peak mass loading is much higher with a value of $\sim 20$ close to the disc and $\sim 1$ at $z=2 \ {\rm kpc}$.  This blows out most of the gas from the disc causing the large drop off in the SF (see Fig.~\ref{fig:sfr}) during the post starburst phase of the simulations.  During this phase of evolution ($t=25-50 \ {\rm Myr}$), $\eta_{\rm M}$ in the ESF runs drops precipitously. This is caused by the large drop off in the star formation by almost a order of magnitude which reduces the pressure behind the outflow, making it stall and fall back onto the disc. Once the disk settles down $\eta_{\rm M}$ rebounds to values of about $2-3$ close to the disc, while there is very little gas at $z \gtrsim 1 \ {\rm kpc}$. We note that this value is higher than that obtained in the SN simulation by a factor of $5-10$. It is quite clear that a true large scale wind is only launched during the initial starburst phase, while only a small scale fountain flow operates after $t>75 \ {\rm Myr}$ in all the runs. The initial starburst in our simulations is caused by the gas radiatively cooling, loosing its pressure support and settling down in the external gravitational potential into a thin disk. This is compounded by the fact that the main channel of feedback in our simulations, SNe, are only active $5 \ {\rm Myr}$ after the first stars form. Therefore, self-regulation during this initial period is not possible. Changes in initial conditions and parameters such as the time period between star formation and SNe feedback will change the duration and strength of the starburst. While, the reasons for a starburst in the simulations are largely dictated by the way the initial conditions of our simulations are constructed, the difference between the evolution of SN and ESF simulations is quite dramatic and deserves further examination. Furthermore, an analogy can be made to
the systems that have undergone a starburst during a merger making
the scenario of a post merger starburst induced quenching in
galaxies more probable.

 \begin{figure*} 
\includegraphics[scale=0.45]{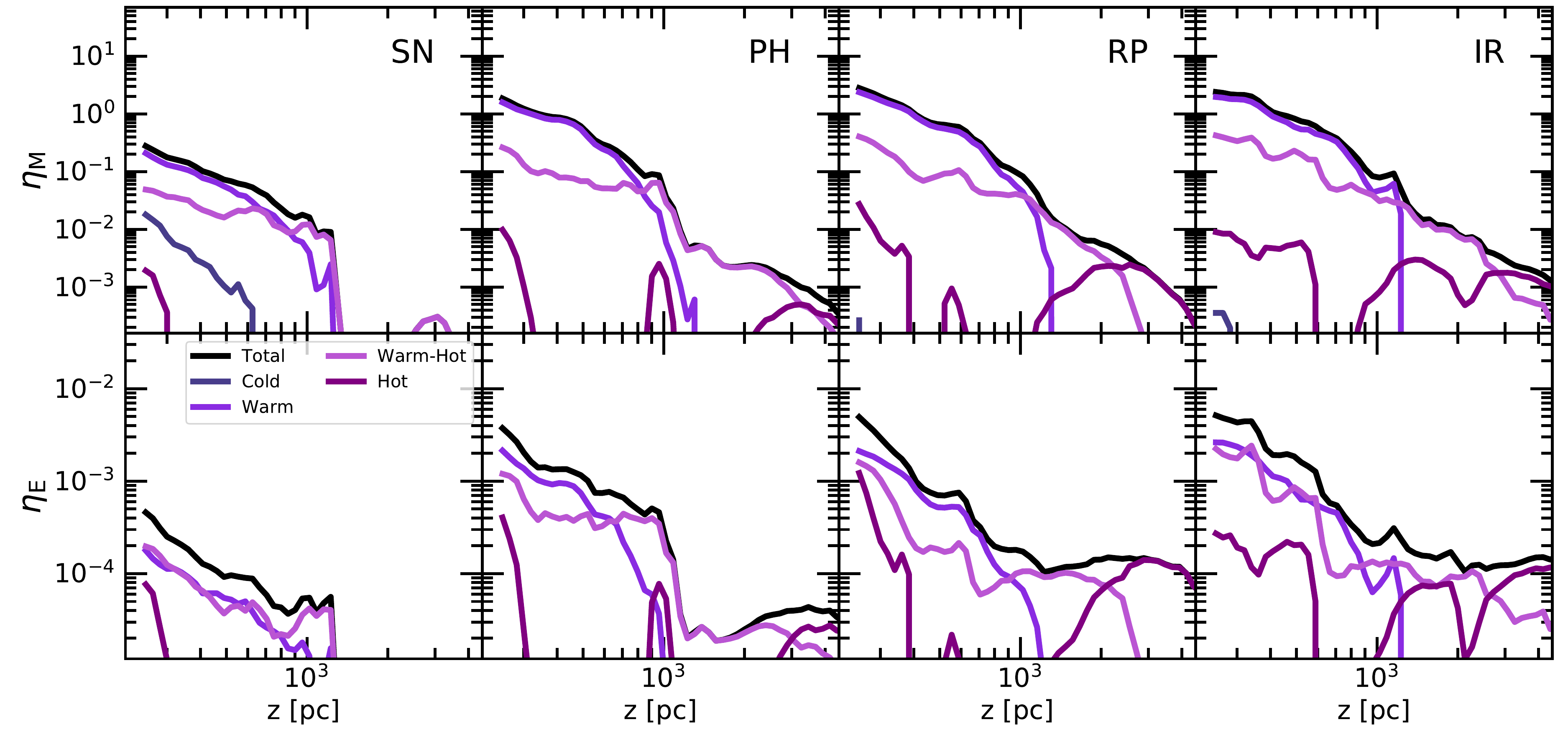}
\caption{The mass (top panles) and energy loading (bottom panels) factors as a function of height from the disc (z) for cold (${\rm T} \leq 300 \ {\rm K}$), warm ($300 < {\rm T} \leq 8000 \ {\rm K}$), warm-hot ($8000 < {\rm T} \leq 3\times 10^5 \ {\rm K}$) and hot gas (${\rm T} > 3\times 10^5 \ {\rm K}$) during the fountain flow phase of the simulation, $t=150$ Myr. The fist column shows the values for the SN simulation, second column; PH, third column; RP and fourth column; IR.  $\eta_{\rm M}$ and $\eta_{\rm E}$ are lower than during the outflow phase of the simulation, but the trends are quite similar.}
\label{fig:dcT150}  
\end{figure*}

\begin{figure*} 
\includegraphics[scale=0.45]{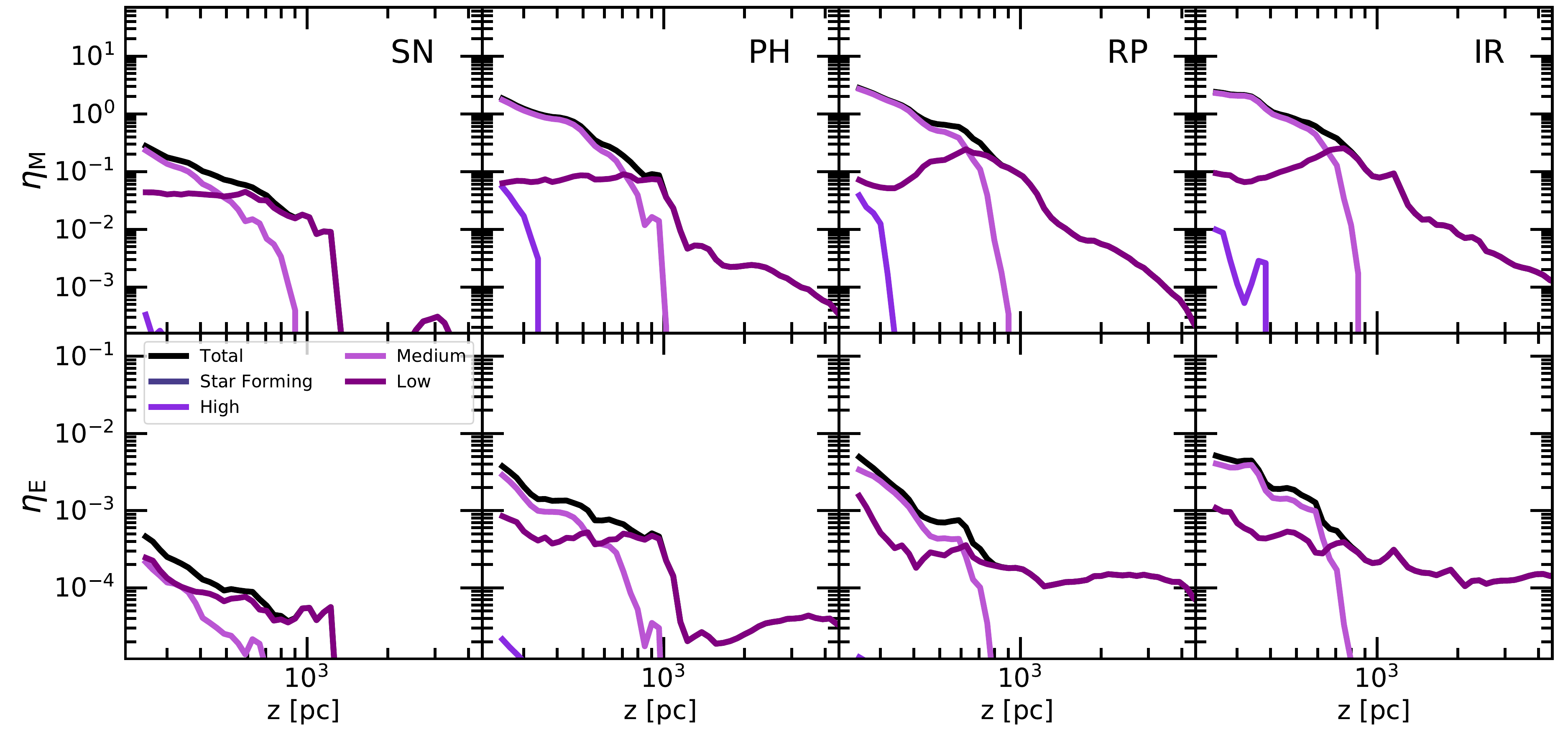}
\caption{The mass (top panels) and energy loading (bottom panels) factors as a function of height from the disc (z) for star forming ($n \geq 10^2 \ {\rm cm}^{-3}$), high ($1\leq n< 10^2  \ {\rm cm}^{-3}$), medium ($10^{-2} \leq n< 1  \ {\rm cm}^{-3}$) and low ($n<10^{-2} \ {\rm cm}^{-3}$) density gas during the fountain flow phase of the simulation, $t=150$ Myr. The fist column shows the values for the SN simulation, second column; PH, third column; RP and fourth column IR.}
\label{fig:dcn150}  
\end{figure*}

 Fig.~\ref{fig:Eload} shows the energy loading factor  at $0.25  \ {\rm kpc}$ (first panel), $0.5  \ {\rm kpc}$ (second panel), $1  \ {\rm kpc}$ (third panel) and $2  \ {\rm kpc}$ (fourth panel) above the disc for the SN (red curves), PH (green curves), RP (green curves) and IR (cyan curves) simulations. The evolution of $\eta_{\rm E}$ is qualitatively quite similar to the evolution in  $\eta_{\rm M}$, with the peak $\eta_{\rm E}$ occurring at about the same time as the occurrence of peak $\eta_{\rm M}$ at $t\sim 30 \ {\rm Myr}$, followed by a precipitous drop in the post starburst period for the ESF runs and the existence of a small-scale fountain flow with low energy loading after $t\gtrsim 75  \ {\rm Myr}$. Peak $\eta_{\rm E}$ in the ESF runs ranges from $\sim 0.02$ at $z = 0.25 \ {\rm kpc}$ to $3 \times 10^{-3}$ at $z = 2 \ {\rm kpc}$. Interestingly, the peak energy loading in the SN simulation remains constant at about $3 \times 10^{-3}$ across different heights from the disc. At late times value of $\eta_{\rm E}$ in the ESF runs is about $2-3$ times larger than in SN run close to the disc ($z\lesssim 1 \ {\rm kpc}$). It is difficult to make any claims about $\eta_{\rm E}$ above this height because of the absence of outflowing material.

 A more comprehensive picture of the outflow behaviour can be obtained by decomposing (Fig.~\ref{fig:dcT30} ) the mass (top panels) and energy (bottom panels) loading into the different temperature bins at the peak of the outflow ($t=30 \ {\rm Myr}$) as a function of height from the disc. The dark blue curves indicate $\eta_{\rm M}$ and $\eta_{\rm E}$ for cold ($T<300 \ {\rm K}$) gas, violet curves for the warm ($300 \ {\rm K}\leq T<8000 \ {\rm K}$) gas, orchid curves for the warm-hot ($8000 \ {\rm K} \leq T< 3 \times 10^5 \ {\rm K}$) gas and purple curves, the hot ($T\geq 3 \times 10^5 \ {\rm K}$) gas. Close to the disc, the SN run has a mass loading of about $\sim 2$ and decreases to $0.02$ by $z \sim 4 {\rm kpc}$, the energy loading remains constant at about $3 \times 10^{-3}$. The ESF runs on the other hand, have $\eta_{\rm M}$ of $\sim 10$ close to the disc which,   reduces to $\sim 0.02$ by about $\sim 4 \ {\rm kpc}$, while $\eta_{\rm E}$ ranges from $0.02$ to $\sim 2 \times 10^{-3}$. Cold gas is never entrained in the outflow irrespective of the simulation we consider. The hot phase in the SN run has slightly higher mass and energy loading compared to the ESF runs. However, both these phases are subdominant to the warm and warm-hot gas phases. The warm-hot gas has roughly a constant mass ($\eta_{\rm M} \sim 0.1$) and energy ($\eta_{\rm E} \sim 2 \times 10^{-3}$) loading up to $z=4 \ {\rm kpc}$, implying the existence of a true large-scale wind \citep{Kim2017}. Including radiation fields only slightly increases $\eta_{\rm M}$ and $\eta_{\rm E}$ of this phase but drastically boosts it for the warm gas, especially at low $z$ ($<2 \ {\rm kpc}$). This effect is so large that the higher mass and energy loading in the ESF runs at low $z$ is almost entirely driven by the increase in the warm phase of the outflow. This implies that the runs with radiation fields are able to entrain more warm material in the outflow and eject it to distances of about $2 \ {\rm kpc}$ from the disc. Therefore, the ESF runs launch a more pronounced small scale, warm fountain flow in addition to the large scale warm-hot wind during the starburst phase. This also explains the almost constant total energy loading in the SN run, contrasted to the declining energy loading in the ESF runs.

\begin{figure*} 
\includegraphics[scale=0.5]{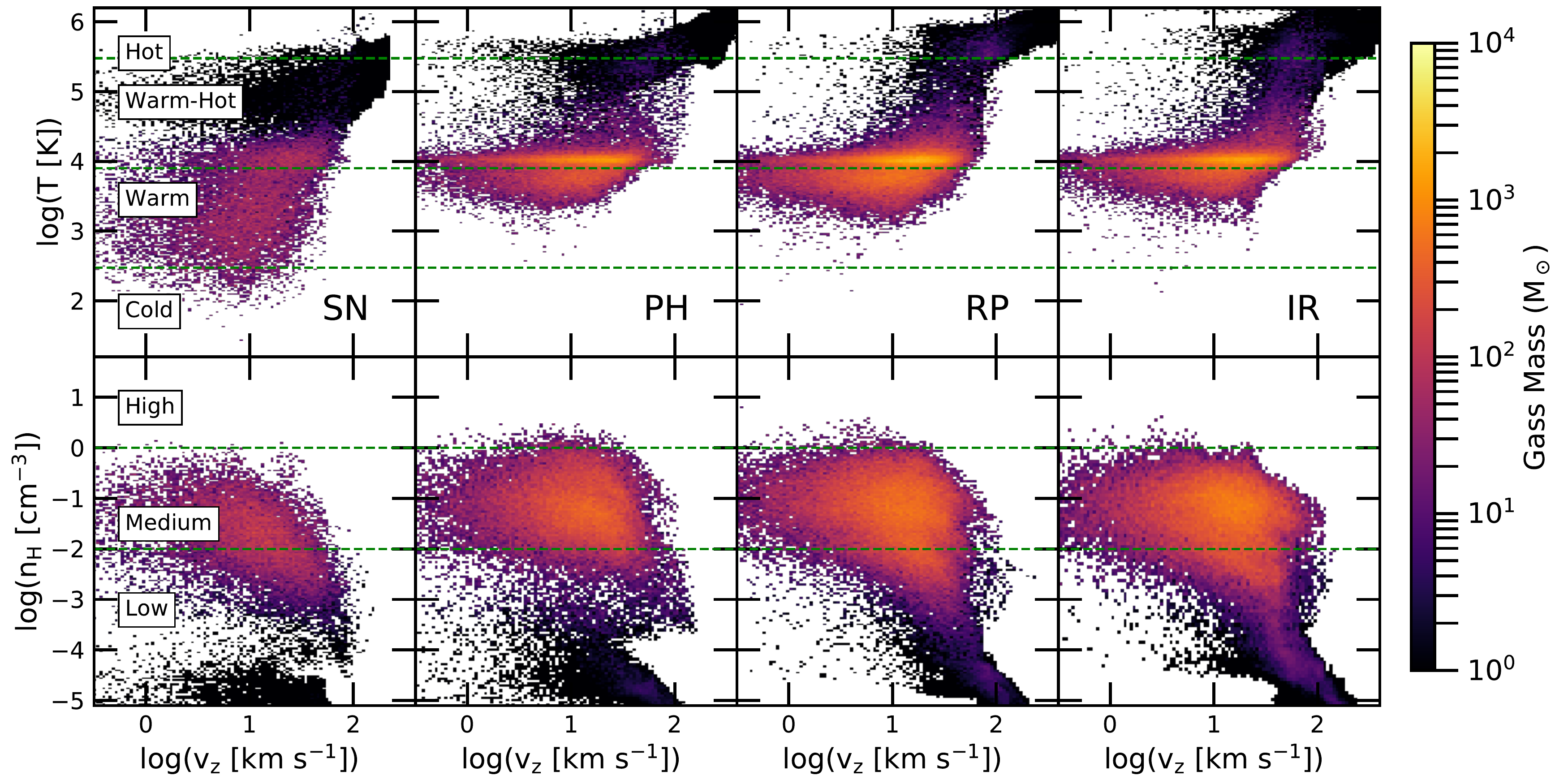}   
\caption{The temperature (top panels) and density (bottom panels) histograms of the outflowing gas as a function of the outflow velocity at $t=150$ Myr for the SN (first column), PH (second column), RP (third column) and IR (fourth column) simulations.}
\label{fig:vel150}  
\end{figure*}
 
A similar analysis can be performed by decomposing the outflow into star forming ($n \geq 10^2 \ {\rm cm}^{-3}$; dark blue curves), high ($1\leq n< 10^2  \ {\rm cm}^{-3}$; violet curves), medium ($10^{-2} \leq n< 1  \ {\rm cm}^{-3}$; orchid curves) and low ($n<10^{-2} \ {\rm cm}^{-3}$; purple curves) density gas (Fig.~\ref{fig:dcn30}).  The density structure of the outflow in many ways mirrors its temperature structure. There is no entrainment of the star forming gas in any of the runs. Some amount of high density gas is launched upto heights of $z\lesssim 0.5  \ {\rm kpc}$ and the  $\eta_{\rm M}$ and $\eta_{\rm E}$ for this phase in the ESF runs is about 10 times larger than in the SN run. The outflow is mostly dominated by the medium and low density material. The medium density gas forms a fountain flow that reaches heights of $z\lesssim 2 \ {\rm kpc}$ while the low density material forms a true large scale wind. The mass and energy loading of the low density wind is quite similar in the SN and ESF runs, but the medium density fountain flow is about $10$ times more mass and energy loaded in the ESF runs compared to the SN run.  Therefore a picture emerges of a starburst driven outflow that can be decomposed into two distinct phases, a small-scale ($\sim 2 \ {\rm kpc}$) fountain flow mainly composed of warm ($300 \leq T < 8000 \ {\rm K}$), medium density ($0.01 \leq n < 1 \ {\rm cm}^{-3}$) gas and a large scale wind ($\gtrsim 4 \ {\rm kpc}$) that is composed of warm-hot ($8000 \leq T < 3 \times 10^5 \ {\rm K}$), low density ($n < 10^{-2} \ {\rm cm}^{-3}$) material. We conclude that radiation fields coupled with SNe have the effect of launching colder, denser and higher mass loaded outflows compared to feedback models that do not consider this process.

In order to better understand the mutiphase, multi-component nature of the outflow we plot (Fig.~\ref{fig:vel30}) the temperature-outflow velocity (top panels) and density-outflow velocity (bottom panels) histograms in the SN (first column), PH (second column), RP (third column) and IR (fourth column) simulations. This plot considers all the gas that is moving outward from the disc and at a height $z>250 \ {\rm pc}$ as part of the outflow.  The green horizontal lines divide the outflow into star forming, high, medium and low density gas and similarly  into cold, warm, warm-hot and hot gas. There is a very clear trend towards higher velocities for higher temperature and lower density gas.  There are some very interesting differences between the SN and ESF runs. The amount of gas mass in the outflow is much larger in the runs with radiation fields. ESF runs are also able to entrain more cold and high density material.  More importantly the warm-medium density gas phase is more mass loaded in the ESF runs.  The velocity of this phase is however less than $\sim 100 \ {\rm km \ s} ^{-1}$, meaning that it cannot get very far from the disc, thereby generating a small scale fountain flow. The material with velocities larger than $100  \ {\rm km \ s} ^{-1}$ go on to generate the large scale wind which is both low density and hot. This explains the mass and energy loading behaviour of the outflows in our simulations. We note that the outflows described here are unlikely to reach wind velocities large enough to be unbound from the galaxy, because the Milky Way escape velocity at the solar circle probably exceeds $500 \ {\rm km \ s}^{-1}$ \citep{Smith2007}.

A similar analysis can be performed in the fountain flow phase of the simulation ($t>75 \ {\rm Myr}$). Fig.~\ref{fig:dcT150} decomposes the mass (top panels) and energy (bottom panels) loading into the different temperature bins during the fountain flow phase of the simulation at $t=150 \ {\rm Myr}$ as a function of height from the disc. The mass and energy loading factors are lower than in the starbusrt phase as expected. Moreover, they decline quite quickly as the height above the disc increases, which is a clear sign of a fountain flow.  Close to the disc, the SN run has a mass loading of about $\sim 0.2$ and decreases to $< 10^{-3}$ by $z \sim 1 {\rm kpc}$, the energy loading goes from $6 \times 10^{-4}$ to $\lesssim 10^{-4}$ in the same range. The ESF runs on the other hand, have $\eta_{\rm M}$ of about $\sim 2$ close to the disc and reduces to $\sim 10^{-3}$ at about $\sim 4 \ {\rm kpc}$ and $\eta_{\rm E}$ ranges from $3 \times 10^{-3}$ to $\sim 5 \times 10^{-4}$. In the SN run, none of the temperature phases reach beyond $z \gtrsim 1 {\rm kpc}$. This is not the case in the runs with radiation fields, where, the warm-hot material reaches heights of $\sim 3 \ {\rm kpc}$ and above this height the hot phase starts to dominate the outflow.  The warm phase dominates both the mass and energy loading of the outflow within about $z = 1 \ {\rm kpc}$ in the ESF runs, but remains sub-dominant in the SN run. 

 \begin{figure*} 
\includegraphics[scale=0.39]{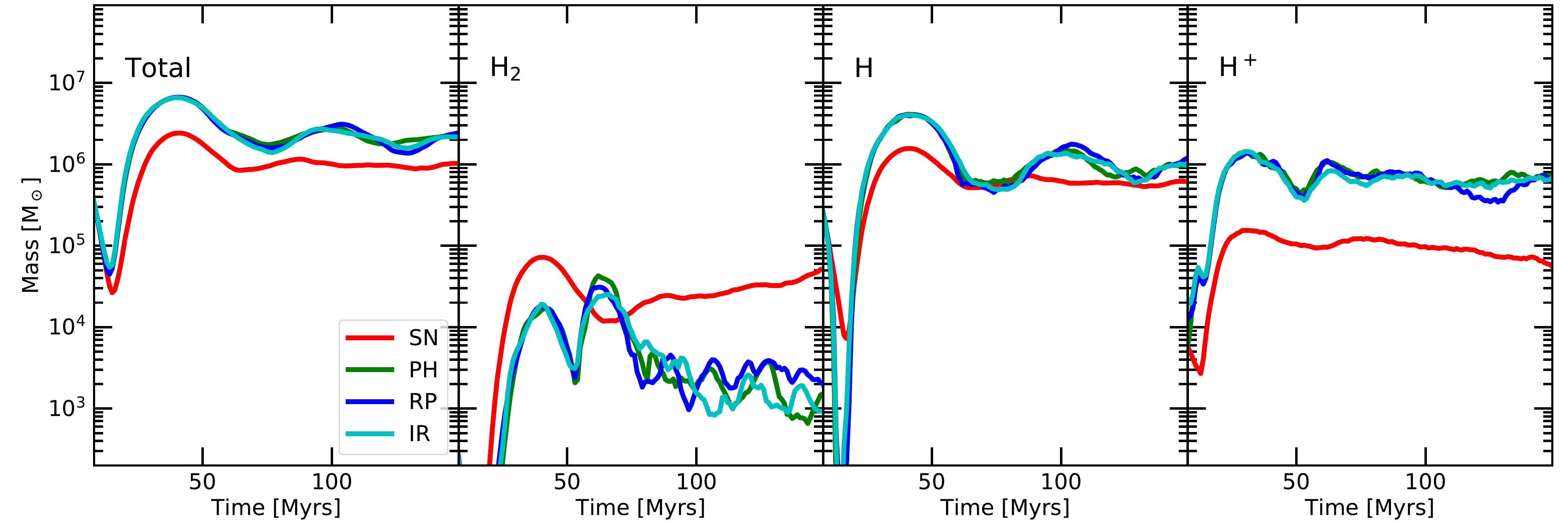}   
\caption{The chemical composition fo the ejected outflow as a function of time in the SN (red curves), PH (green curves), RP (blue curves) and IR (cyan curves) simulations. The ejected material is defined as any material with ${\rm z}>250 \ {\rm pc}$.  The first column shows the total amount of ejected gas, the second column -- mass in molecular hydrogen (${\rm H}_2$), third column -- neutral hydrogen and the fourth column -- mass of ionized hydrogen.} 
\label{fig:massH}
\end{figure*}

Fig.~\ref{fig:dcn150} decomposes the outflow into the star forming, high, medium and low density gas (Fig.~\ref{fig:dcn30}) at $t=150 \ {\rm Myr}$.  Similar to the starburst phase, the density structure of the fountain flow in many ways mirrors its temperature structure.  There is no entrainment of the star forming gas in any of the runs. The SN run manages to entrain very little high density material but the ESF runs can entrain and launch this material up to $z\leq 0.3 \ {\rm kpc}$.  Most of the mass and energy loading in the ESF runs at low $z$ ($\lesssim 1 \ {\rm kpc}$) is dominated by the medium density gas, whereas, in the SN run, the low density gas dominates at all heights except in the very center. There is no appreciable gas above $z>1 \ {\rm kpc}$ in the SN run, but there is some in the ESF runs, with most of the gas at this height composed of low density-hot material. The additional entrainment of the medium density gas increases the mass and energy loading in the ESF runs by a factor of $\sim 5-10$.

Fig.~\ref{fig:vel150} shows the the temperature-outflow velocity (top panels) and density-outflow velocity (bottom panels) histograms in the SN (first column), PH (second column), RP (third column) and IR (fourth column) simulations at $t=150 \ {\rm Myr}$. All gas that is moving outward from the disc and at a height $z>250 \ {\rm pc}$ is considered a part of the outflow. The green horizontal lines divide the outflow into star forming, high, medium and low density gas and similarly  into cold, warm, warm-hot and hot gas.  The amount of mass in the outflow is much lower than during the  starburst phase. The cold high density phase no longer exists. There is very little gas with velocities $v_z \gtrsim 100 \ {\rm km \ s}^{-1}$, explaining the lack of a large scale wind.  In the runs with radiation fields, most of the gas is photoionized and has a temperature of $\sim 10^4 \ {\rm K}$. The medium density - warm-hot phase is more mass loaded in the ESF runs compared to the SN run. It is therefore, quite clear that even in the low star formation mode of evolution, radiation fields make stellar feedback more efficient. They do not generate large scale winds, but are able to launch low temperature, higher density outflows that are higher mass and energy loaded. These outflows reach heights of about $1-2 \ {\rm kpc}$ from the disc, forming a robust fountain flow. In contrast just SNe feedback creates a weak fountain flow that consists of mainly very low density hot gas. This is despite the fact that the ESF runs have about two times lower SFR than the SN run. Importantly,  $\eta_{\rm M}$ and $\eta_{\rm E}$ are about a factor of $\sim 5-10$ larger in the presence of radiation fields and this increase is mainly driven by the entrainment of warm, dense material in the outflowing gas. This quantitative result is holds true in both the starbusrt and fountain flow phases of the simulation therefore implying that this is a very robust prediction.

 \begin{figure*} 
\includegraphics[scale=0.33]{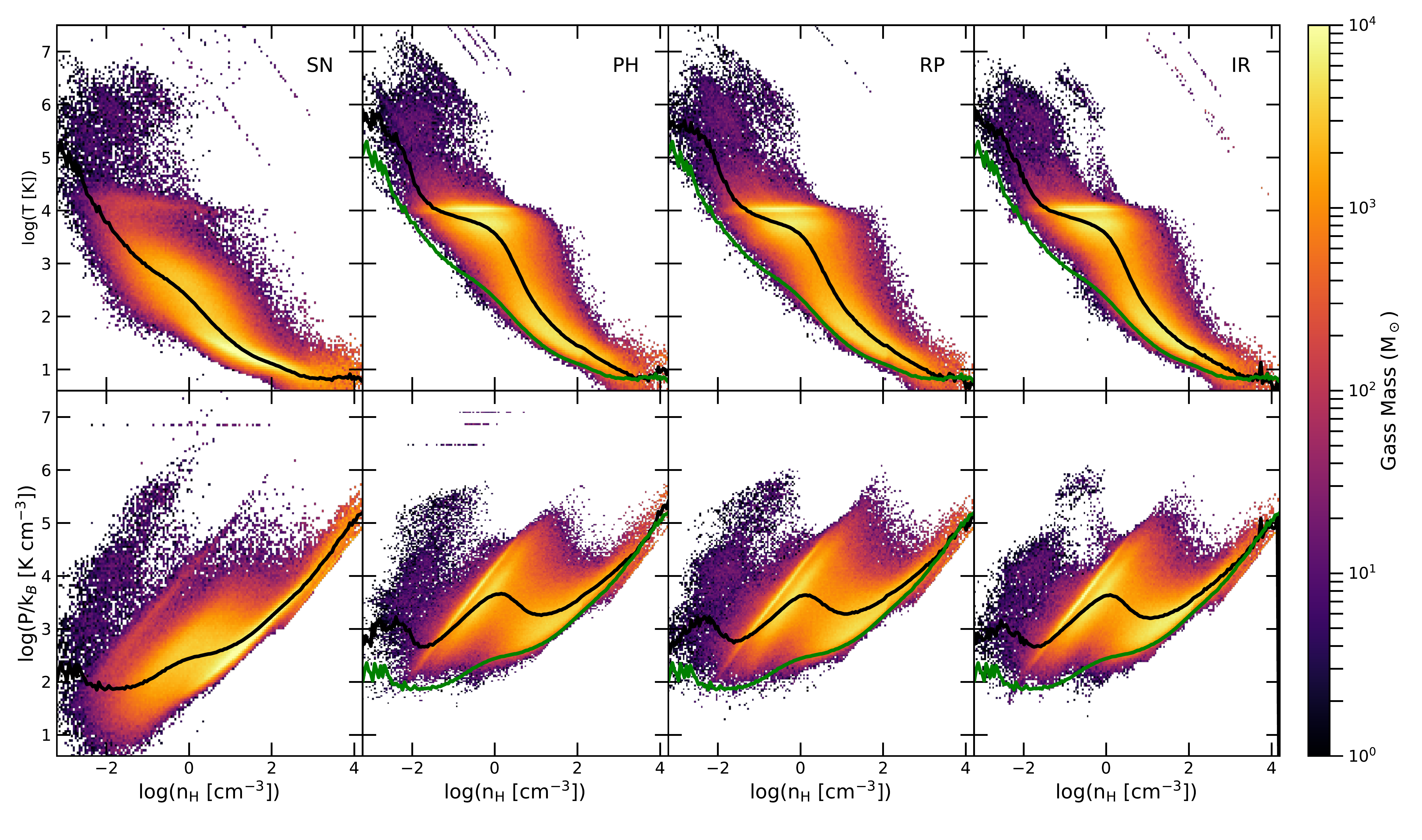}   
\caption{The temperature-density (top panels) and pressure-density (bottom panels) phase space diagram of the disc (${\rm z}<250 \ {\rm pc}$) at $t=150$ Myr for the SN (first column), PH (second column), RP (third column) and IR (fourth column) simulations. In general for a given density the gas temperature and therefore pressure is higher for the runs with ESF. The black lines denote the median of the distribution and the solid green lines in the ESF runs denote the median obtained from the SN simulation.} 
\label{fig:phasespace}
\end{figure*}

To assess the impact of radiation fields on the chemical composition of the outflowing gas, we plot in Fig.~\ref{fig:massH} the masses of the ejected material in terms of different hydrogen species as a fucntion of time in the SN (red curves), PH (green curves), RP (blue curves) and IR (cyan curves) simulations. The ejected material is defined as any gas present above $z=250 \ {\rm pc}$ at any given point in time.  The first column shows the total amount of ejected gas, the second column -- mass in molecular hydrogen ${\rm H}_2$, third column -- neutral hydrogen and the fourth colum -- mass of ionized hydrogen. As expected, the total amount of gas ejected from the disc is higher in the ESF runs, by about a factor of $\sim 5$ during the starburst period and about $\sim 2$ at the later stages ($t\gtrsim 75 \ {\rm Myr}$). The fraction of gas in the molecular state is quite low ($<5\%$) in all the runs. This is expected because the density of gas at these heights is quite low $<1 \ {\rm cm}^{-3}$. However, the lack of any radiation fields in the SN run allows for a comparatively large molecular fraction in the outflow. This picture will definitely change if the metagalactic background is included. The larger amount of cold, dense gas in the ESF outflows, does allow for self-shielding of the hydrogen molecule giving rise to a larger fraction of neutral hydrogen gas in the outflow. Finally, we note that the illumination of the ejected gas by the radiation fields from the disc increases the abundance of ionized hydrogen by almost a factor of $\sim 10$. Therefore, we conclude that the ESF runs give rise to a ejecta that is half-ionized and half neutral, whereas the ejecta in the SN run is mainly neutral due to the absence of radiation fields.

\section{Discussion}
\label{sec:discussion}
On the whole the runs with radiation fields increase the mass and energy loading factor by about a factor of $5-10$, close to the disc ($z<1 \ {\rm kpc}$), throughout the entirety of the simulation time. They are still lower by about $2-3$ when compared to random driving (SNe explosions occur at random positions) tall box simulations presented in \citet{Girichidis2015}. The motivation of these simulations was to get the structure of the ISM right, meaning that the mass and energy loading factors were not tuned to match observations. On the other hand our values are closer to the those predicted by  global galaxy simulations with more realistic geometries \citep{Muratov2015, Fielding2017} and with analytic estimates of the values required to explain the observed galaxy stellar mass function and the metal enrichment of the intergalactic medium \citep{Somerville2015}.  However, our tallbox geometry implies that the wind mass and energy loading factors are not well-defined because they decline significantly with increasing box height (See Section~\ref{sec:conclusions} for more details). Despite this caveat, it is quite clear that early stellar feedback does indeed play an                 important role in regulating star formation and launching outflows.  The consistently lower SFRs in the ESF runs indicate that radiation fields make feedback more efficient. There are two ways in which the radiation fields can reduce star formation. They can heat up the disc through photoheating, which increases the temperature in the disc, puffing it up and reducing the star formation.  Secondly, photoheating and radiation pressure can evacuate gas from the neighbourhood of newly formed stars allowing them to explode in relatively low density environments. This increases the momentum output of SNe explosions by reducing the cooling loses \citep{Martizzi2015}. There is evidence for both these effects in our simulations.

\begin{figure} 
\includegraphics[width=\columnwidth]{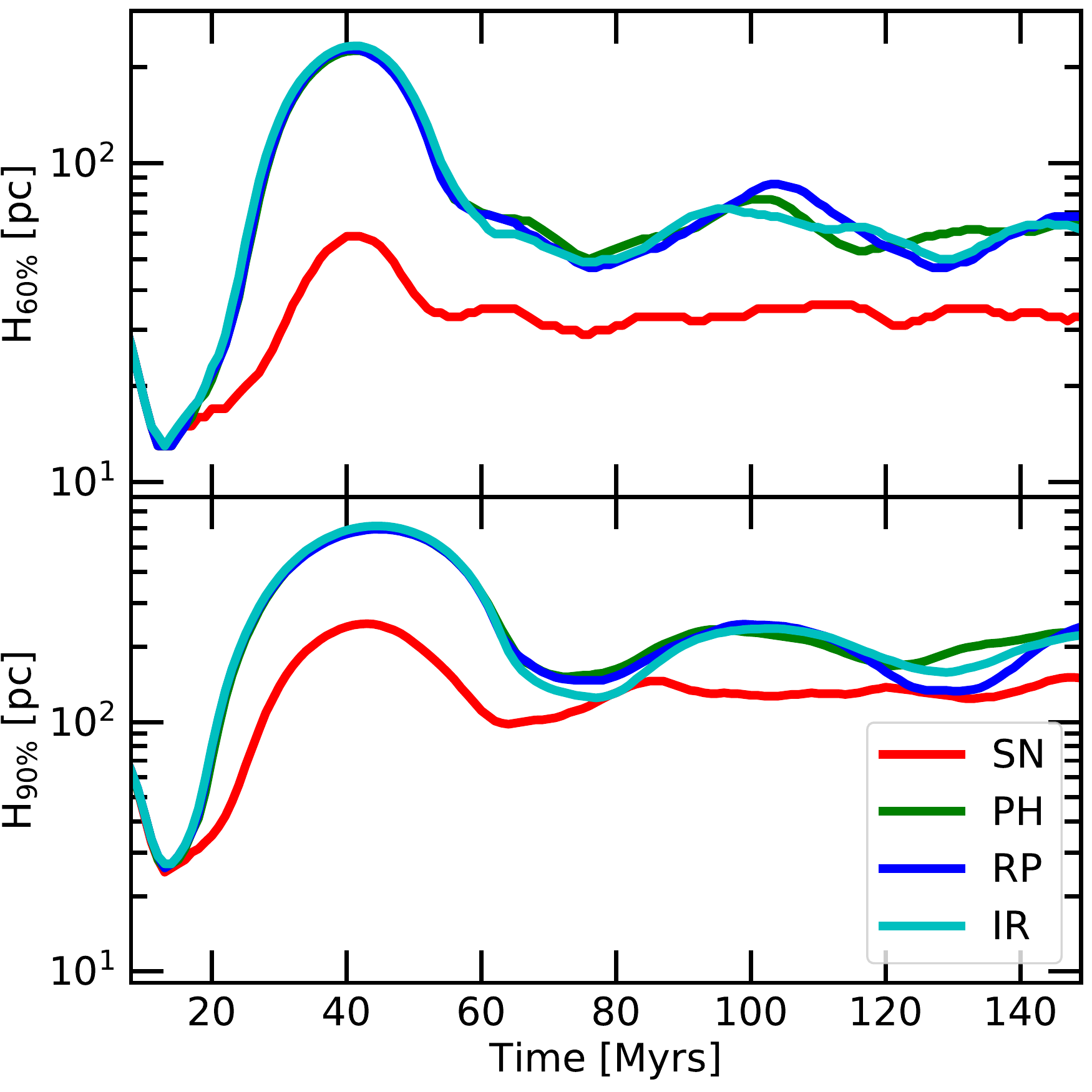}  
\caption{Vertical heights of $60\%$ (top panel) and $90\%$ enclosed mass as a function of time for the SN (red curves), PH (green curves), RP (blue curves) and IR (cyan curves) simulations. ESF increases the scale height of the disc almost a factor of $\sim 2$. } 
\label{fig:height}  
\end{figure}

Fig.~\ref{fig:phasespace} shows the temperature-density (top panels) and pressure-density (bottom panels) phase space diagram of the gas in the disc (defined as $z<250 \ {\rm pc}$) at $t=150 \ {\rm Myr}$ in the SN (first column), PH (second column), RP (third column) and IR (fourth column) simulations. The black solid lines show the median of the distribution. For comparison,  the median obtained from the SN simulation is plotted in the ESF runs in solid green lines. Only the ESF runs show a true two-temperature multiphase medium in pressure equilibrium. For a  given density the temperature and therefore the pressure of the gas is higher in the runs with radiation fields. This higher pressure provides additional support against gravitational collapse, puffing up the disc and reducing the star formation rate. This effect can be quantified by looking at the vertical structure of the disc.  A simple estimate is obtained by calculating the height which encloses $60\%$ of the total mass of the disc (${\rm H_{60\%}}$) and contrasting it with ${\rm H_{90\%}}$, the height which encloses $90\%$ of the total mass. ${\rm H_{60\%}}$ gives us an estimate of the distribution of the dense molecular gas in the disc, while ${\rm H_{90\%}}$ informs us about the envelope of the disc. Fig.~\ref{fig:height} shows the time evolution of ${\rm H_{60\%}}$ (top panel) and ${\rm H_{90\%}}$ (bottom panel) for all the simulations. For the initial $\sim 10 \ {\rm Myr}$, the disc cools and contracts reducing ${\rm H_{60\%}}$ to $\sim 10 \ {\rm pc} $ and ${\rm H_{90\%}}$ to $\sim 20 \ {\rm pc}$. This gas compression leads to a starburst-induced outflow leading to an increase in both ${\rm H_{60\%}}$ and ${\rm H_{90\%}}$. The stronger outflows in the ESF runs cause ${\rm H_{90\%}}$ to reach a peak of about $\sim 600 \ {\rm pc}$ at $t=40 \ {\rm Myr}$, while it only increases to about $\sim 200 \ {\rm pc}$ in the SN run. At the same time ${\rm H_{60\%}}$ increases to $\sim 200 \ {\rm pc}$ in the ESF runs and $\sim 60 \ {\rm pc}$ in the SN run. Both heights decrease during the inflow period and then rebound back and remain fairly constant after $t> 75 \ {\rm Myr}$. At these late times, the difference in ${\rm H_{90\%}}$ between the SN ($\sim 150 \ {\rm pc}$) and ESF runs is quite nominal ($\sim 200 \ {\rm pc}$), with the ESF runs having a larger value by about $50 \%$. ${\rm H_{60\%}}$, on the other hand shows about a factor of two increase in the ESF  ($\sim 60 \ {\rm pc}$) runs. This implies that the discs in ESF runs are more puffed up due to additional pressure from the photoheated gas, which in turn reduces the midplane pressure reducing the SFR. 

\begin{figure}  
\includegraphics[width=\columnwidth]{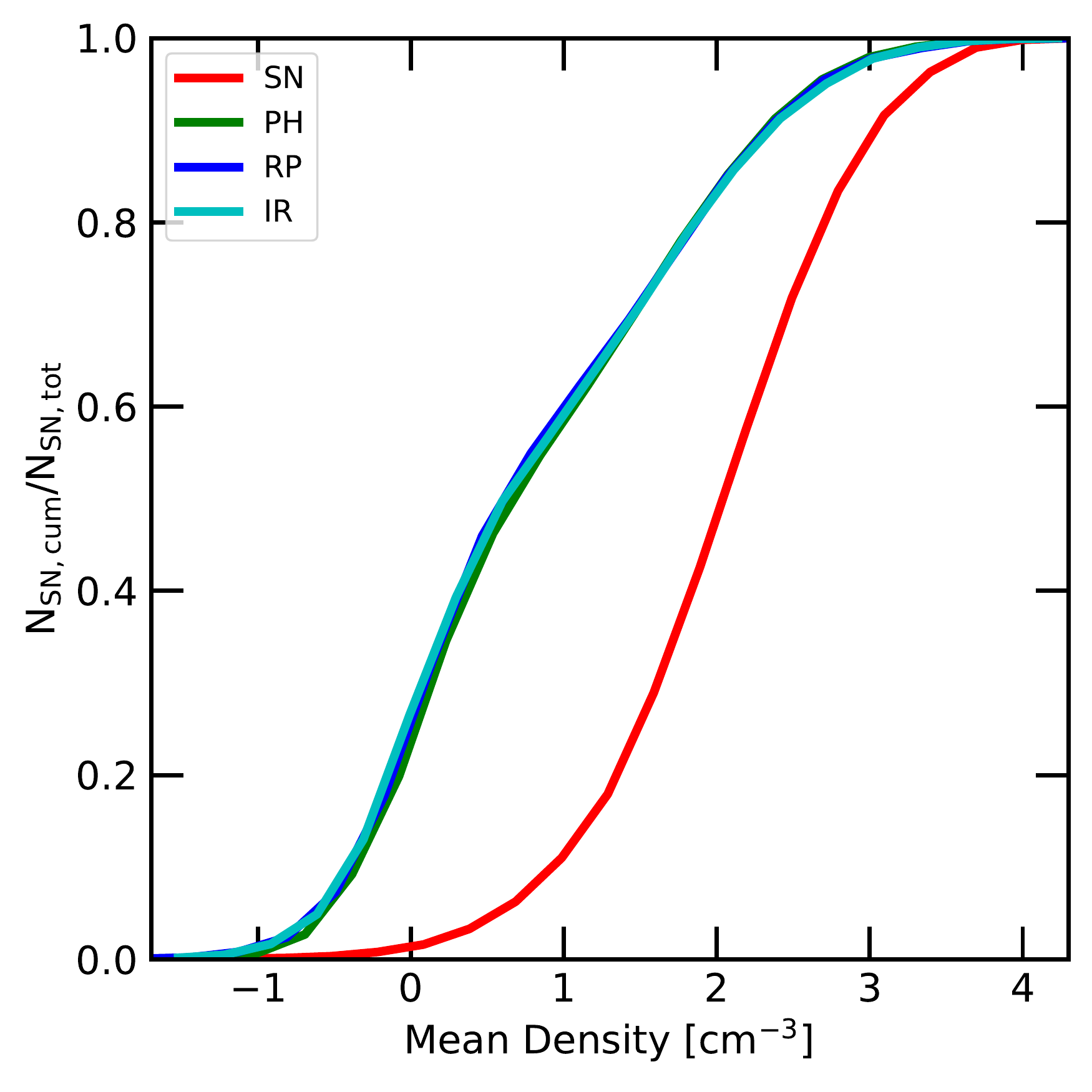} 
\caption{Normalised cumulative counts of the supernovae as a function of the mean environmental density in which they explode for the SN (red curves), PH (green curves), RP (blue curves) and IR (cyan curves) simulations. Photoheating reduces the density at which the SNe go off by a factor of $10-100$, thereby increasing the momentum input of a SNe feedback event.}  
\label{fig:meandens}  
\end{figure}

It is also worth noting that the temperature and pressure difference between the SN and ESF runs in the high density star forming gas ($n_{\rm H} \gtrsim 100 \ {\rm cm}^{-3}$) is quite small, implying very little difference between the sites of star formation in the different runs. However, once the stars form, photoheating of high density material around a newly formed star, over-pressurises the region, which then expands till it reaches a pressure equilibrium with the surrounding gas. This reduces the density of the gas and causes a pileup of material at the photeheating temperature ( about $\sim 10^4 \ {\rm K}$ for a soft spectra from stars), a prominent feature in the ESF runs. Direct evidence for this effect can be gained by looking at the mean densities at which the SNe explode in the different runs. Fig.~\ref{fig:meandens} plots the cumulative distribution of the number of SNe as a function of the mean density at which they explode in the SN (red curve), PH (green curve), RP (blue curve) and IR (cyan curve) simulations.  In SN run, about $50\%$ of the SNe explode in star forming high density gas and about $90\%$ in gas with $n > 10 \ {\rm cm}^{-3}$. Since the star formation density threshold is $n = 100 \ {\rm cm}^{-3}$, this means that the SNe in the SN run are rarely able to get out of their respective birth clouds. On the other hand, in the ESF runs, the more than $80\%$ of the SNe explode in gas with $n < 10 \ {\rm cm}^{-3}$ and more than $90\%$ in gas with $n < 100 \ {\rm cm}^{-3}$, implying that radiation fields reduce the mean ambient density by a factor of about $10-100$.

The evolution of a SNe remnant in different density environments have been studied in great detail by many recent works \citep{Kim2015, Martizzi2015, Haid2016}. Briefly, the evolution of the SNe can be divided into five phases \citep{McKee1977}, the pre Sedov-Taylor (PST) phase, the Sedov-Taylor (ST) phase, the transition phase (TR), the pressure driven snowplough (PDS) phase and finally the momentum conserving (MC) phase. During the PST phase, the SNe ejecta runs into the ambient ISM shocking and thermalizing a large fraction of the injected energy. This then initiates a energy conserving ST phase, which ends when the rate-of-change of temperature due to adiabatic expansion is comparable to radiative losses. The expansion then transitions to a regime, where the adiabatic and radiative cooling contributes about equally called the transition phase. Finally, the PDS phase is defined by the domination of radiative cooling with the pressure inside the bubble driving the expansion of the SNe remnant. During all these evolutionary phases the incresead thermal pressure within the bubble pushes on the surrounding low pressure ISM boosting the final momentum output of the SNe event. Eventually, when the pressure inside the bubble becomes equal to the ambient pressure, the remnant becomes momentum-conserving. Simulations have shown that the momentum boost achieved in realistic ISM environments is about a factor of $13-30$ depending on the ambient density in which the SNe explodes.  The duration of the ST, TR and PDS phases depends on the cooling time of the gas, which in turn depends on the ambient density, with more cooling occurring in higher density gas ($t_{\rm cool} \propto n^{-0.55}$). The longer cooling times in low density medium leads to a longer duration of the momentum boosting phases leading to more ISM material being swept up and accelerated to larger velocities, with the mass of the swept up material scaling with ambient density as $M_{\rm swept} \propto n^{-0.26}$ which implies that a reduction in the ambient density by a factor of $\sim 100$ increases the mass of the swept material by about a factor of $\sim 3$. Therefore, we can conclude that because SNe explode in lower density environments in the ESF simulations, the duration of the momentum boosting phases is longer, leading to a higher amount of the colder, denser ISM (compared to the hot wind) being swept up and accelerated to larger velocities in the outflow. This not only explains the increased mass and energy loading factors in the ESF runs, but also the increased entrainment of colder and denser gas.

Finally, we note that the resolved SNe momentum budget has only a weak dependence on the density of the ambient medium $\dot{p} \propto n^{-0.19}$ \citep{Kim2015, Martizzi2015, Haid2016}. Using this scaling we can conclude that radiation fields can increase the momentum output of the SNe by a factor of about $1.5-2.5$.  This increased coupling efficiency combined with the fact that the radiation fields provide additional pressure support against gravity explains the differences between the SN and ESF runs presented in this paper. These effects are in agreement with previous self-consistent RHD simulations presented in \citet{Rosdahl2015} $\&$ \citet{Peters2017}.

\section{Conclusions}
\label{sec:conclusions}

In this paper, we present extremely high resolution radiation hydrodynamic simulations of a small patch of the inter-stellar medium performed with {\sc Arepo-RT}. We performed four simulations, SN: simulation with just SNe feedback, PH: SN + photoheating from local radiation fields, RP: PH + effect of UV radiation pressure and IR: RP + effect of multiscattering infrared radiation pressure. These simulations were performed with a mass resolution of $10 \ {\rm M}_\odot$ and a spatial resolution of $\sim 45 \ {\rm pc}$. This allows us to resolve all the relevant feedback processes, thereby providing us with an accurate picture of stellar feedback in low gas surface density galaxies. Our main results can be summarised as follows:
\begin{itemize}

 \item Radiation fields have the effect of reducing the star formation rates and thereby the total stellar mass of the galaxies by about a factor of $\sim 2$. This has the effect of increasing the gas depletion timescales in the simulations, thereby allowing for a better match with the observed Kennicutt-Schmidt relation. 
 
 \item The most important effect of radiation fields is photheating through UV photons. Radiation pressure both single and multiscattered, does not have a significant effect in low gas surface density environments.

  \item The mass and energy loading factors increase by $\sim 5-10$ in the presence of radiation fields. The increase is mainly driven by the additional entertainment of medium density ($10^{-2} \leq n< 1  \ {\rm cm}^{-3}$), warm ($300 \ {\rm K}\leq T<8000 \ {\rm K}$) material in the outflow. This material has velocities of about $\sim 10-100 \ {\rm km  \ s}^{-1}$, meaning that it falls back onto the disc creating a fountain flow of order $\sim 2  \ {\rm kpc}$. Radiation fields, therefore, help drive colder, denser and higher mass and energy  loaded outflows compared to models that invoke only SNe feedback.  
  
  \item The radiation fields from stars generates an inter-stellar radiation field (ISRF) that permeates through the disc increasing the temperature of the gas which in turn increases the pressure support of the gas against gravitational collapse. This puffs up the disc and reduces star formation.
  
  \item Photoheating of high density material around a newly formed star over-pressurises the region, which then expands till reaches a pressure equilibrium with the surrounding gas. This reduces the ambient density in which the SNe explodes by a factor of $10-100$, increasing its momentum input by a factor of $\sim 1.5-2.5$. 
\end{itemize}

The relatively high star formation rates even with radiation fields hint that there might be additional feedback mechanisms that have not been accounted for in our present work such as stellar winds \citep{Gatto2017, Peters2017}, binary stars \citep{Kim2017} or cosmic rays \citep{Simpson2016, Ruszkowski2017, Diesing2018}. We plan to include the effect of all three processes in a future work.

The effect of radiation pressure is minimal when it acts together with photoheating, because the latter is substantially more efficient in inducing velocities
comparable to the sound speed of the hot ionized medium on time-scales far
shorter than required for accumulating similar momentum with radiation pressure. This allows photoheating to dominate the feedback as the heating and expansion of gas
lowers the central densities, further diminishing the impact of radiation pressure \citep{Sales2014}. In the future, we plan to build upon this work and test the effect of radiation pressure in high surface density, star-bursting galaxies \citep{Kleinmann1988} as well as in more massive giant molecular clouds in low surface density galaxies, as this mechanism is theorised to be more effective in these environments \citep{Hopkins2011, Thompson2015, Rahner2017}.

Another caveat of this work is that the geometry of the tallbox setup is not realistic. Previous works have shown that the properties of galactic winds are not accurately predicted because they do not capture the correct global geometry and gravitational potential of galaxies \citep{Martizzi2016}. The wind mass and energy loading factors are not well-defined because they decline significantly with increasing box height. However, this local tallbox setup was required to achieve the resolution necessary to resolve relevant feedback processes in our simulations. We plan to run physically realistic calculations (e.g., isolated full disc simulation) to fully understand the role of radiation fields in launching galactic outflows in the future. While, we have only focused on the dynamical impact of early stellar feedback, radiation fields also change the chemical composition of the ISM which we will investigate in a future paper. We conclude by noting that early stellar feedback in the form of photoheating is an important physical process that enhances the effectiveness of SNe feedback, a result that is in agreement previous self-consistant RHD simulations \citep{Rosdahl2015, Gatto2017, Peters2017, Emerick2018}, thereby confirming the important role that radiation fields play in regulating star formation and determining the structure of the ISM and outflows in galaxies.

\section*{Acknowledgements}
RK acknowledges support from NASA  through Einstein Postdoctoral Fellowship grant  number PF7-180163 awarded by the \textit{Chandra} X-ray Center, which is operated by the Smithsonian Astrophysical Observatory for NASA under contract NAS8-03060. SCOG acknowledges support from the Deutsche Forschungsgemeinschaft (DFG) via SFB 881 ``The Milky Way System'' (sub-projects B1, B2 and B8).  Computing resources supporting this work were provided by the NASA High-End Computing (HEC) Program through the NASA Advanced Supercomputing (NAS) Division at Ames Research Center and by XSEDE project AST180025 via the Comet supercomputer center at San Diego.

\bibliographystyle{mnras}
\bibliography{paper} 

Kleinmann1988
\appendix

\section{\ion{H}{$_2$} thermochemistry}
\label{sec:h2}

 \begin{figure}  
\includegraphics[width=\columnwidth]{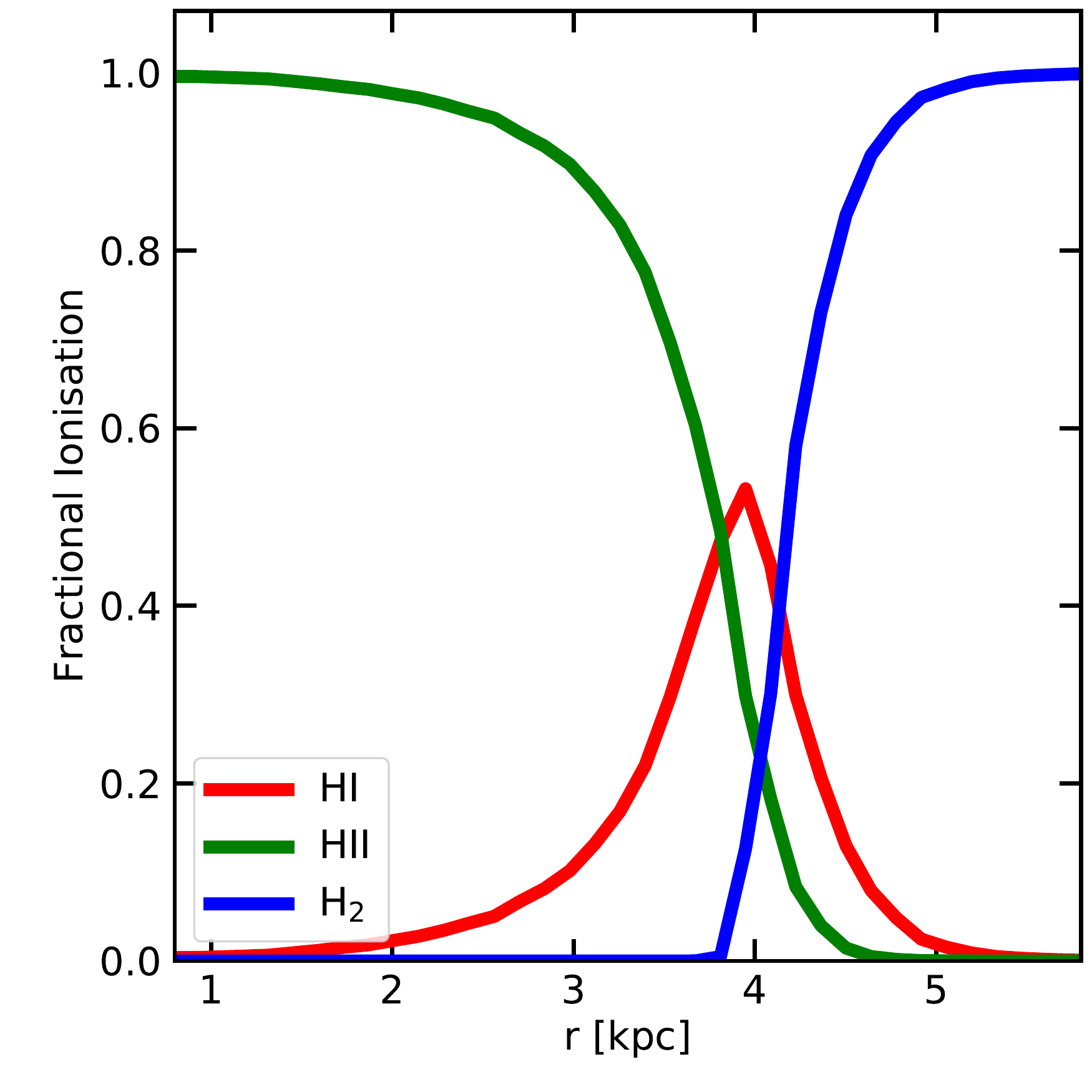} 
\caption{Profiles of the neutral hydrogen (red curve), ionized hydrogen (green curve) and molecular hydrogen (blue curve) fractions as a function of radius around an O star at the end of the molecular Str\"omgren sphere test.}  
\label{fig:h2}  
\end{figure}

As a test for our \ion{H}{$_2$}, thermochemistry we perform a Str\"omgren sphere test in a molecular medium \citep[][Section 3.4]{Nickerson2018}. We start with a completely molecular solar-metallicity gas at a density of $n_H = 10^{-3} \ {\rm cm}^{-3}$ and a temperature of $13.56 \times 10^3 \ {\rm K}$. The simulation box has a side of length $15 \ {\rm kpc}$ which is resolved by $32^3$ resolution elements. We do not include any photoheating or cooling effects. The central ionization source is a O star which emits radiation spectrum equivalent to a blackbody with an effective temperature of $4.3 \times 10^4 \ {\rm K}$. Three frequency bins used in this test are, the Lyman-Werner (LW; \ion{H}{$_2$} dissociation) band ($11.2-13.6\ {\rm eV}$), hydrogen ionization band   ($13.6-15.2\ {\rm eV}$) and \ion{H}{I} and \ion{H}{$_2$} ionization band  ($15.2-24.6\ {\rm eV}$). The total luminosity is set such that the number of \ion{H}{$_2$} dissociation photons $\dot{N}_{H_2} = 3 \times 10^{48} \ {\rm s}^{-1}$ and the number of \ion{H}{I} ionization photons is $\dot{N}_{HI} = 5 \times 10^{48} \ {\rm s}^{-1}$.  The  \ion{H}{$_2$} ionization rates are taken from \citet{Baczynski2015} and the \ion{H}{$_2$} dissociation rates and the self-shielding prescriptions are obtained from \citet{Nickerson2018}. For these conditions the radius of the Str\"omgren sphere is
\begin{equation}
 r_{s, {\rm \small{HI}}} = \left(\frac{3\dot{N}_{\rm \small{HI}}}{4\pi \alpha_{\rm \small{HI}} n_H^2}\right)^{1/3} \sim 4.1 \ {\rm kpc} \ ,
\end{equation}
where $\alpha_{\rm \small{HI}}$ is the Case B recombination rate. 

Fig.~\ref{fig:h2} shows the fractional ionization profiles of molecular (blue curve), neutral (red curve) and ionized (green curve) hydrogen after $500 \ {\rm Myr}$ of evolution.  The \ion{H}{II}
region ends sharply at $4.1 \ {\rm kpc}$ as expected. The self-shielding prescription is able to block most of the LW photons from entering into the molecular layer prducing a relatively sharp and thin \ion{H}{I} layer between the fully ionized  \ion{H}{II} and fully molecular \ion{H}{$_2$} layers. This nicely matches with the results obtained by \citet{Nickerson2018}, thereby confirming the accuracy of our scheme.

\bsp
\label{lastpage}
\end{document}